\documentclass[pre,amsmath,twocolumn,showpacs,superscriptaddress]{revtex4-1}
\usepackage{graphicx}
\usepackage{color}
\usepackage{latexsym}
\usepackage{amssymb}
\usepackage{longtable}
\usepackage{subfigure}

\begin{document}

\title{Bacterial range expansions on a growing front:\\
Roughness, Fixation, and Directed Percolation}

\author{Jordan M. Horowitz}
\affiliation{Physics of Living Systems Group, Department of Physics, Massachusetts Institute of Technology, 400 Technology Square, Cambridge, MA 02139}
\affiliation{Department of Biophysics, University of Michigan, Ann Arbor, Michigan, 48109,
USA}
\affiliation{Center for the Study of Complex Systems, University of Michigan, Ann Arbor, Michigan 48104, USA}
\author{Mehran Kardar}
\affiliation{Department of Physics, Massachusetts Institute of Technology, Cambridge, Massachusetts 02139, USA}

\date{\today}

\begin{abstract}   
{\it Directed Percolation} (DP) is a classic model for nonequilibrium phase transitions into a single absorbing state
({\it fixation}).
It has been extensively studied by analytical and numerical techniques in diverse contexts. 
Recently, DP has appeared as a generic model for the evolutionary/ecological dynamics of competing bacterial populations. {\it Range expansion} -- 
the stochastic reproduction of bacteria competing for space to be occupied by their progeny --
 leads to a fluctuating and rough {\it growth front}, which is known from  experiment and simulation to affect the underlying critical behavior of the DP transition.  In this work, we employ symmetry arguments to construct a pair of non-linear 
stochastic partial differential  equations describing the co-evolution of surface roughness with the composition field of DP.
Macroscopic manifestations (phenomenology) of these equations on growth patterns and genealogical tracks of range
expansion are discussed; followed by a renormalization group analysis of possible scaling behaviors at the DP transition. 
\end{abstract}

\maketitle 

\section{Introduction}

\begin{figure}[tb]
\includegraphics[scale=.29]{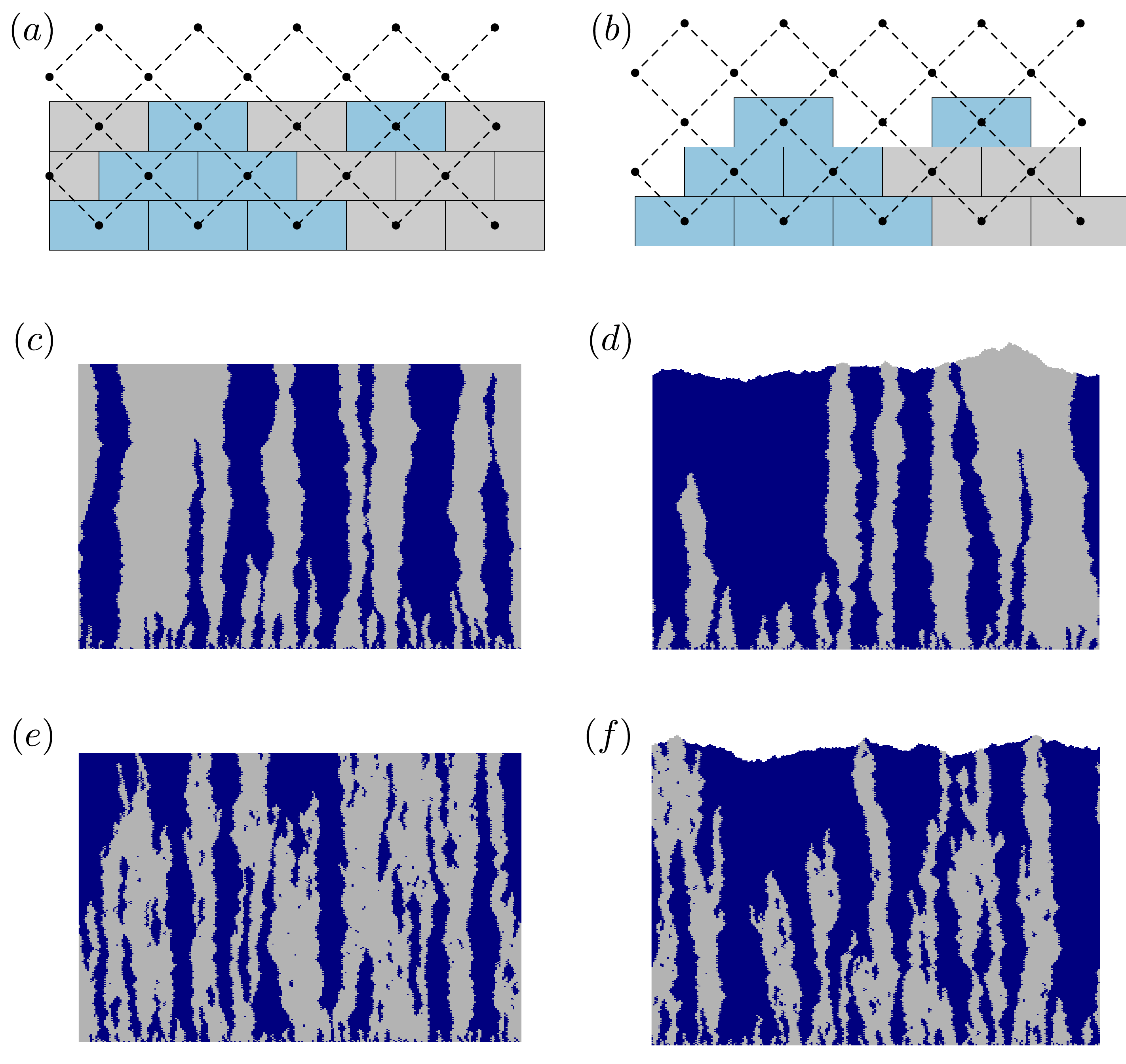}
\caption{Numerical illustration of DP with rough surface growth: Active particles (or wild-type bacteria) are depicted in gray with inactive sites (or mutant bacteria) in blue.  Growth proceeds vertically with a flat front grown one layer at a time $(a)$ or a rough front $(b)$ with each new addition's identity determined by its local neighborhood. In $(c)$ and $(d)$ the two species are identical leading to compact regions of growth. In $(e)$ and $(f)$ we allow for mutations that can appear in the active wild-type regions, but not vice versa, making the lack of active wild-type sites an absorbing state.  
}\label{fig:illustration}
\end{figure}

Directed percolation (DP) serves as the paradigmatic example of a nonequilibrium  phase transition between 
an active state (of coexisting components) and an absorbing state (with one component)~\cite{Henkel,Odor2004,Tauber}.
Quantitative understanding of the critical behavior at this transition is rooted in numerous theoretical~\cite{Janssen2005,Tauber2005} and numerical~\cite{Hinrichsen2000,Henkel} analyses.    
As depicted in Fig.~\ref{fig:illustration}, DP can be visualized as the irreversible and directed growth of active sites on a lattice, with new sites added to the growing cluster one layer at a time.
However, it is its interpretation as a single-species reaction-diffusion process that offers its most comprehensive formulation and has lead to the suggestion of its widespread prominance in the form of the {\em DP hypothesis}: the critical properties near an active-to-absorbing phase transition should generically be governed by the DP scaling exponents, provided the stochastic process is Markovian, the order parameter is decoupled from any other slow variable, there is no quenched disorder and there are no special symmetries~\cite{Henkel,Tauber2005,Tauber}.

Recently, a model of competition between bacteria in an expanding population has been proposed as a new manifestation of the DP universality class~\cite{Korolev2010,Korolev2011}.
Such bacterial range expansions, however, differ from traditional DP in one important aspect:
the growing population front can be rough, an effect we highlight in Fig.~\ref{fig:illustration} with sample realizations of a toy model.
It is now widely appreciated that such a rough frontier can have significant influence on the evolutionary/ecological dynamics~\cite{Hallatshcek2007,Hallatschek2010,Korolev2012,Farrell2017}.
Experimental evidence reinforced by simple scaling arguments further suggests that the influence of rough growth leads to inter-species domain boundary fluctuations not consistent with the DP universality class~\cite{Hallatshcek2007}.
This analysis has been corroborated by a handful of numerical simulations of various modifications of the two-species Eden model allowing for cooperation, competition and/or mutations~\cite{Saito1995,Kuhr2011,Lavrentovich2014,Kuhr2015}.
All have confirmed that the scaling exponents of the absorbing phase transition differ from those at the DP transition.
In addition, when the two species have different expansion rates, the measured scaling exponents for the frontier fluctuations appear distinct from those predicted by the Kardar-Parisi-Zhang (KPZ) equation -- a classic minimal model of surface growth~\cite{Kardar1986}.

In this article, we complement these studies with a field-theoretic analysis based on symmetry considerations.
The resulting phenomenological model is composed of a pair of  stochastic partial differential equations
that describe the coupled time evolutions of the growth front and the DP composition field.
Our template for this approach are previous studies in Refs.~\cite{Drossel2000,Drossel2003}, where the effects of surface roughening on phase ordering were modeled by coupling an Ising order parameter to the KPZ equation for surface growth, as well as earlier studies of coupled KPZ equations~\cite{EK93}.
The construction of the model is detailed in Sec.~\ref{sec:model}, where we start by recounting field theoretic
formulations of the DP transition, and KPZ growth, and then indicate how different types of symmetry
allow and constrain couplings between these fields.
The phenomenology of the minimal set of couplings is discussed in Sec.~\ref{sec:manifestations}.
In particular, we elucidate the macroscopic manifestations of the couplings on the shape of the growth
fronts, and on the dependence of various velocities and transition points on the overall slope of the front.
Section~\ref{sec:maps} discusses hidden symmetries of the model, and its mapping to other problems of
interest. In particular, it is shown that in certain limits the deterministic version of the equations is exactly
solvable (following linearization by a Cole-Hopf transformation), highlighting the emergence of 
fixation starting from a rough initial condition.
The renormalization group analysis, carried out in Sec.~\ref{sec:RG},
 then reveals that the coupling between the DP and KPZ fields is relevant, 
modifying the scaling exponents of both  the DP transition and the KPZ equation.
Previous renormalization group calculations in this direction considered the effects of surface fluctuations with correlated spatial noise on the absorbing phase transition, but ignored feedback from the DP order parameter onto the surface~\cite{Sarkar2012}, which we explicitly include.

\section{Coupling composition and roughness at a front}\label{sec:model}

The DP active-to-absorbing phase transition can be analyzed in the continuum limit by considering
the variations of a field $f({\bf x},t)$ that classically represents the density of active particles at time $t$ and position 
${\bf x}$ in a $d$-dimensional space.
It is then the disappearance of active particles due to death or coalescence that signals the transition 
to the absorbing state, $f=0$.
Here, our motivation comes from range expansions of immotile bacteria~\cite{Hallatshcek2007}, where  we  
instead consider the competition between a pair of bacterial species: a wild type and a mutant.
In this case, the order parameter $f$ characterizes the fraction of wild-type bacteria, with the transition to the absorbing state mediated by the irreversible takeover of the mutants.
From this point of view, $f$ reflects the composition of the growing bacterial colony.

In this section, we introduce various limiting dynamics for this bacterial composition  field and its coupling to surface fluctuations, working up stepwise from simpler to more complex, terminating with the full coupled roughness/DP equations.
The goal is to pinpoint the new phenomenology introduced with each new level of complexity.
To aid in the discussion, it is helpful to have in mind a representative microscopic dynamics, like the ones pictured in Figs.~\ref{fig:illustration}$(a)$ and $(b)$. In the evolutionary biology community, the underlying ``stepping stone'' model was introduced by Kimura and Weiss~\cite{Kimura1964} and has been used extensively since then to examine competitive growth~\cite{Korolev2010, Korolev2010,Lavrentovich2013}.
The model is equivalent to the  Domany-Kinzel (DK) cellular automaton rules for flat growth~\cite{Korolev2010,Lavrentovich2013}, and the rough variant is based on the ``brick wall'' solid-on-solid model used previously for the study of phase transitions in the growth of binary films~\cite{Drossel2000,Drossel2003}, or to the ``single step'' growth model introduced by Meakin \emph{et al.\ }\cite{Meakin1986} and Plischke \emph{et al.\ }\cite{Plischke1987}.
We will discuss their specifics in more detail in Sec.~\ref{sec:manifestations}, but, briefly, different species -- wild-type (gray bricks) and mutant (blue bricks) -- grow out on a  tilted square lattice by replicating into adjacent empty lattice sites with rates that can depend on their local neighborhood~\cite{Korolev2010}.
The growth can  be either restricted to the addition of one full new layer at a time [Fig.~\ref{fig:illustration}$(a)$], or growth can evolve organically allowing for a roughening of the surface~[Fig.~\ref{fig:illustration}$(b)$].

\subsection{Flat  growth without mutation}

The simplest variant of the model starts with a mixture of two species that grow (possibly at different rates)
one layer at a time.
In the absence of mutations, the distinction between wild-type and mutant disappears, and 
either species can potentially take over the population (\emph{i.e.}~there are two  absorbing states).  
Figure~\ref{fig:illustration}$(a)$ (in the absence of mutation) suggests a mean-field dynamics (ignoring fluctuations) where the fraction of the first species $f$ grows in proportion to the difference in growth rates $\tau^\prime$ as $\partial_t f=\tau^\prime f(1-f)$.
Allowing for demographic noise and migration (diffusion) leads to a Langevin equation,
describing the evolution of the field $f({\bf x},t)$ as~\cite{Korolev2010,Janssen2005b}
\begin{equation}\label{eq:CDP}
\partial_t f = K\nabla^2 f+\tau^\prime f (1-f)+\sqrt{2D_f f(1-f)}\eta.
\end{equation}
Here, $K$ is the spatial diffusion (or migration) rate, and $\eta({\bf x},t)$ is delta-correlated (in space and time) noise, with intensity (variance) $D_f$.

Equation~\eqref{eq:CDP} describes a critical, albeit discontinuous, phase transition as $\tau^\prime$ goes through zero~\cite{Henkel}.
Indeed, for $\tau^\prime<0$ the mean-field steady-state value of the order parameter is $ f^*=0$; whereas, for $\tau^\prime>0$, we have $ f^*=1$.
Thus, the order parameter experiences a discontinuous jump at $\tau^\prime=0$, from one absorbing phase ($ f^*=0$) to another ($ f^*=1$).
The choice of absorbing phase at $\tau^\prime=0$ is a symmetry-breaking consequence of noise
and initial conditions.
This indistinguishability of the two species is reflected in Eq.~\eqref{eq:CDP} by the symmetry $f\to1-f$ and $\tau^\prime\to-\tau^\prime$: interchanging species labels is equivalent to flipping the relative growth rate.
Models that display this qualitative behavior fall into the Compact Directed Percolation (CDP) universality class, which is a special limit of DP~\cite{Henkel}; the nomenclature originates in the observation that the domains of growth are compact, as evidenced in Fig.~\ref{fig:illustration}$(c)$.
Furthermore, as a result, the domain boundaries in a one-dimensional realization, undergo simple diffusion.

\subsection{Flat  growth with mutations}

Adding mutations to the model changes the phenomenology markedly.
Mutations break the symmetry between the species, allowing the mutants to appear within compact regions of the wild-type, as seen in Fig.~\ref{fig:illustration}$(e)$.
In this case, the irreversible takeover of mutants is an absorbing state in which the wild-type cannot invade.
There is now a transition between an active phase with wild-type bacteria, and an absorbing state 
without wild-type bacteria, at the point when the relative growth rate of mutants is balanced by the mutation rate.
Near this critical point the fraction of wild-type bacteria will be small $f\approx 0$, justifying a
small $f$ expansion to capture the long-time, large-lengths critical dynamics with the effective Langevin equation~\cite{Henkel,Tauber2005,Tauber}
\begin{equation}\label{eq:DP}
\partial_t f = K\nabla^2 f+\tau f -\frac{g}{2}f^2+\sqrt{2D_f f}\eta.
\end{equation}
The active-to-absorbing phase transition is now controlled by the parameter $\tau$, which quantifies the difference in growth and mutation rates. 
Indeed, within the mean-field approximation, $\tau<0$ leads to a steady-state solution in the absorbing state $f^*=0$, which is continuously connected to an
active phase when $\tau>0$ with steady-state value $f^*=\tau/2g$, whose magnitude is capped through $g$.
As a minimal model for an absorbing phase transition, Eq.~\eqref{eq:DP} is the field-theoretic Langevin equation capturing the critical behavior of the DP universality class~\cite{Henkel}.
It is worth noting that enforcing $g=\tau$, does not reimpose the CDP symmetry, and thus some care is required in interpreting the connection between universality classes at this level of description.

The DP hypothesis suggests that the critical behavior embodied by Eq.~\eqref{eq:DP} is quite general.
However, by allowing rough growth, thereby coupling the order parameter to the slow dynamics of the surface, we expect novel critical behavior to emerge.

\subsection{Reflection asymmetry in rough front growth}
Since the bacteria reproduce stochastically (not simultaneously in discrete generations) the growth front becomes
rough. Ignoring overhangs and islands, the height of the front is capured by a single-valued height function
$h({\bf x},t)$. {\it For a single bacterial species} the variations in the height of
the front can then be expressed, in the spirit of a gradient expansion, as
\begin{equation}\label{eq:KPZ}
\partial_t h=v_0+\nu\nabla^2h+\frac{\lambda}{2}(\nabla h)^2+\cdots+\sqrt{2D_h}\xi .
\end{equation}
The first term corresponds to the uniform advance of the flat front, the last term $\xi({\bf x},t)$ is
uncorrelated (in space and time) noise that captures the stochasticity of the growth steps, while the intermediate
terms describe the dependence of smoothening and roughening  on the local curvature and the slope of a rough front.

 In the context of interface roughening, distinct scaling behaviors and universality classes are
associated with the cases of $\lambda=0$ (Edwards-Wilkinson~\cite{EW82}) and  $\lambda\neq 0$ (KPZ universality).
The distinction is enforced by an underlying symmetry, in that fluctuations (in the moving frame with $v_0=0$)
are symmetric under reflection $h\to -h$ for $\lambda=0$, with finite $\lambda$ breaking this symmetry.
Thus, at this point we distinguish between two types of
growth that we designate as {\it reflection symmetric}  or {\it reflection asymmetric}.
One way to distinguish between the two classes is to consider growth on an initially sloped
surface with $\nabla h=\sigma$; Eq.~\eqref{eq:KPZ} then predicts an average growth velocity 
$v({\bf \sigma})=\langle \partial_t h\rangle= v_0+\lambda \sigma^2/2 +\cdots$.
If such average for the bacterial film is independent of slope $\sigma$, then $\lambda$
(and higher order terms depending on slope) must be zero. 

Note that Eq.~(\ref{eq:KPZ}) describes the height of the film measured from a flat surface.
Thus if the film grows with the same speed $v_0$ in all directions (\emph{i.e.}\ independently
of slope $\sigma$), the component along the ``vertical" direction increases as 
$v_0\sqrt{1+\sigma^2}\approx v_0+v_0\sigma^2/2+\cdots$.
This simple geometric argument indicates that isotropic growth corresponds to $\lambda=v_0>0$.
However, it is quite possible that due to the manner
by which nutrients are supplied the growth velocity is not isotropic with a different dependence on $\sigma$.
Indeed, for some growth models (see below) $\lambda$ can even have a negative sign.

\subsection{ Profile dependent competition without feedback}

We next extend the above description of rough front growth to the case of wild-type and mutant bacterial
species. 
As a first step in understanding the emergent behavior, we make the simplifying assumption that the 
two species are symmetric with respect to their affect on the front, {\em i.e.},
variations in  height of the film are still described by Eq.~(\ref{eq:KPZ}) with no feedback from $f({\bf x},t)$.
However, we allow variations of the height profile $h({\bf x},t)$ to influence the bacterial field $f$,
which again (in the limit of small $f$ with a gradient expansion) leads to the coupled equations
\begin{align}
\partial_t h&=\nu\nabla^2h+\frac{\lambda}{2}(\nabla h)^2+\sqrt{2D_h}\xi,\\
\nonumber
\partial_t f &= K\nabla^2 f+\tau f -\frac{g}{2}f^2 +\sqrt{2D_f f}\eta\\
&\qquad+\beta_1\nabla f\cdot \nabla h +\beta_2 f \nabla^2 h +\frac{c}{2}f(\nabla h)^2.
\end{align}
Note that, with primary focus on overall shapes and fluctuations, we have set $v_0=0$. 
The effect of height fluctuations on the field $f$ is now captured by the three terms
proportional to $\beta_1$, $\beta_2$, and $c$ whose interpretation will be explained in Sec.~\ref{sec:manifestations}. 
We emphasize that these are the only (possibly relevant) terms appearing at the lowest order in a gradient expansion.
Their effect is to feed the surface fluctuations into the DP equation as an additional source of correlated  (multiplicative) noise.
A comparable set of equations was recently analyzed in order to understand how environmental fluctuations (here surface fluctuations) modify the DP transition, except the authors took $\beta_2=c=0$ and allowed for correlated noise in the KPZ equation, which lead to a somewhat distinct class of behaviors~\cite{Sarkar2012}.
With finite $\beta_1$ and $c$, evolution of the bacterial population is on average dependent on the slope
of the initial front. We designate this case as  {\it profile dependent competition}.

\subsection{ Profile dependent competition with feedback}

Lastly, if the two species expand into the front at different rates as they grow, we  
expect the surface dynamics to be altered by the bacterial composition  field.
Indeed, in models of range expansions of immotile bacteria, differential expansion velocities find their origin in differing growth rates of the bacterial populations~\cite{Kuhr2011,Farrell2017}: whenever bacteria reproduce they replicate into nearby open space, thereby locally advancing the frontier.
Thus, a selective growth advantage ($\tau\neq0$) inherently couples the field $f$ to the surface height.
Allowing for this feedback leads to our most general model of the rough DP transition,
 \begin{align}\label{eq:langevin1}
\partial_t h &= \nu \nabla^2 h +\frac{\lambda}{2} (\nabla h)^2 + \sqrt{2D_h}\xi +\alpha f ,\\ \nonumber \label{eq:langevin2}
\partial_t f &= K\nabla^2 f+\tau f-\frac{g}{2}f^2 +\sqrt{2D_f f}\eta \\ 
&\qquad+\beta_1\nabla f\cdot\nabla h +\beta_2 f \nabla^2 h +\frac{c}{2}f(\nabla h)^2.
\end{align}
This full set of couplings is expected to lead to novel scaling behaviors.
To the best of these authors' knowledge this is the first time such a phenomenological model has been proposed.

In the following, we shall distinguish cases with ($\alpha\neq0$) and without ($\alpha=0$) feedback;
 with profile independent ($\beta_1=\beta_2=c=0$) or dependent ($\beta_1$, $\beta_2$, $c$~ non-zero) competition;
and with ($\lambda\neq0$) or without ($\lambda=0$) front reflection symmetry.
Each combination may in principle belong to a distinct universality classes.
 Note that full reflection symmetry $h\to-h$ in the coupled equations would require $\beta_1=\beta_2=0$, while $c$ may still be
non-zero.

\section{Macroscopic manifestations of coupling parameters}\label{sec:manifestations}

In this section, we discuss and illustrate the macroscopic manifestations of  the phenomenological 
parameters introduced to describe coupling of bacterial  composition to the surface shape.
 While some results of this section, such as the dependence of surface morphology on $\lambda$,
 are well known in the surface-growth community,  
 they may be less familiar in the biological range-expansion circle, and are therefore included for completeness.
The numerical illustrations in this section were generated using a modification of the Domany-Kinzel (DK) cellular automaton~\cite{Korolev2010,Lavrentovich2013} (modified ``brick wall'' solid-on-solid model~\cite{Drossel2000,Drossel2003}, or ``single step'' model~\cite{Meakin1986, Plischke1987}.), whose elementary steps are illustrated in Fig.~\ref{fig:illustration}$(b)$.
Growth proceeds on a lattice of transverse extension $L$ with periodic boundary conditions.
New sites are added randomly as long as the neighboring sites in the layer below are occupied to prevent overhangs.
An added site is assigned one of two `colors' based on the colors of its
local neighborhood through three microscopic parameters: Briefly, they are $\tau_\mu$ which biases the identity of newly added sites based on the selective differences of the colors in the layer below, $\mu_\mu$ a microscopic mutation rate, $\alpha_\mu$ which biases the height growth rate based on the composition field, and $\beta_\mu$ which couples the slope to the composition field by allowing the active neighbors on the same layer to affect the identity of the newly added site.
The role and implementation of these (microscopic) parameters 
is described in more detail in the following paragraphs with additional specifics in Appendix~\ref{sec:Algorithm}.

\subsection{Dependence of front velocity on slope}

\begin{figure}
\centering
\includegraphics[scale=.6]{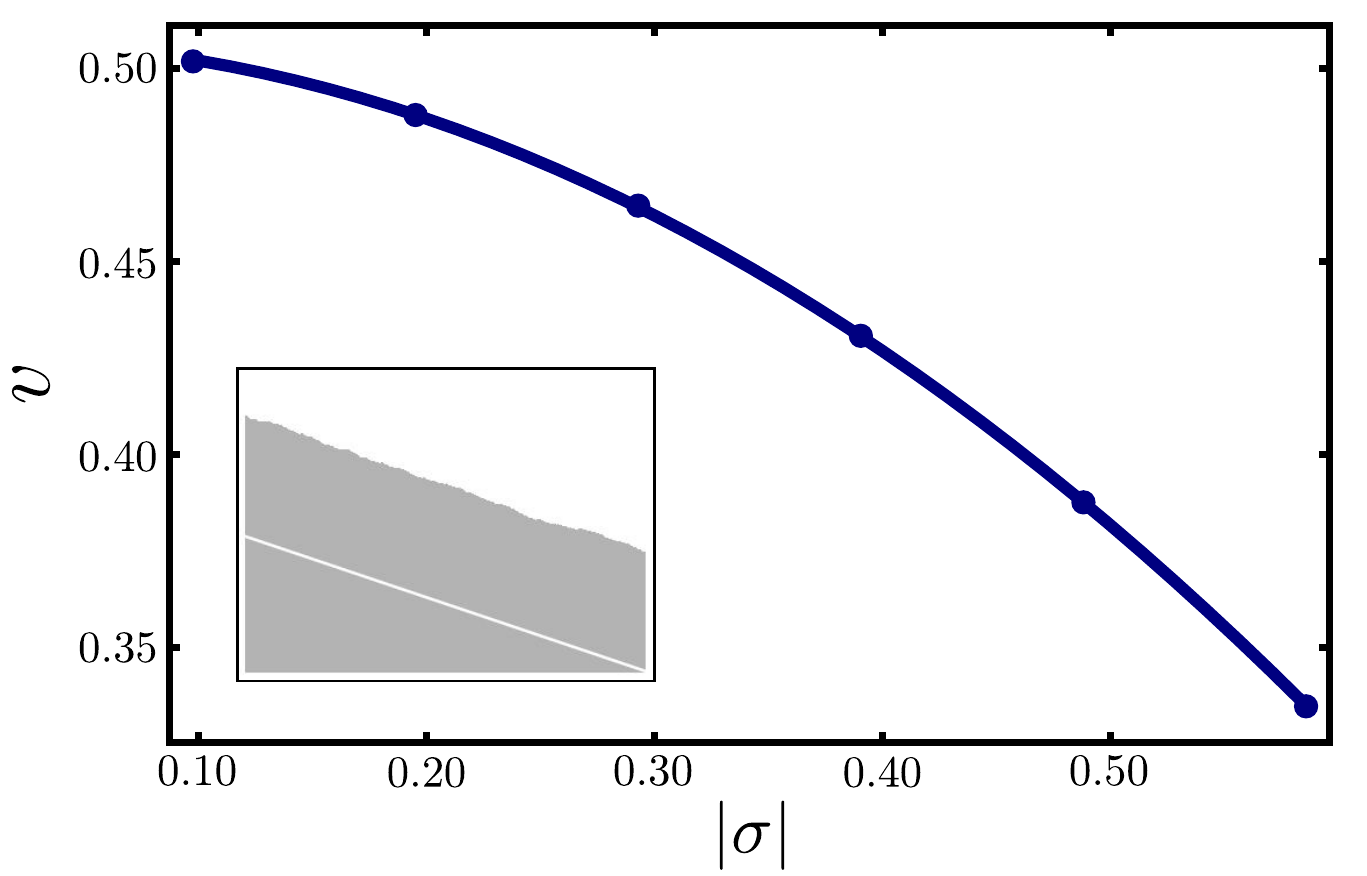}
\caption{Manifestation of the KPZ nonlinearity on growth velocity:  The average front velocity $v$ (in lattice spacings/step) as a function of slope $\sigma$ is fit to $v=v_0+\lambda \sigma^2/2$ with $v_0\approx 0.51$ and $\lambda\approx-1.00$, which agrees well with the exact prediction $v=(1-\sigma^2)/2$~\cite{Krug1992}. Error bars are smaller than data points. Inset: Example of a simulation of the modified DK algorithm for one species with an initial slope $\sigma=-0.6$ (demarcated by white line) and helical boundary conditions.  The evolution maintains a constant slope as predicted by the KPZ equation.  Simulations carried out on a lattice  with $L=256$ and $\alpha_\mu=\beta_\mu=\tau_\mu= \mu_\mu=0$.}\label{fig:lambda}
\end{figure}

In our model, occupation of new  sites is only possible when both neighbors in the layer below are occupied.
This causes growth to be slower on sloped fronts where it is harder to find occupied neighbors, 
suggesting $\lambda <0$ for this model~\cite{Drossel2003}.
 Indeed, for the equivalent ``single step'' model~\cite{Meakin1986, Plischke1987} an exact result by Krug \emph{et al.}~\cite{Krug1992} indicates that $v=(1-\sigma^2)/2$ for such discrete models.
The KPZ equation  maintains the average slope of the front: 
If the initial profile has slope $\nabla h(x,t)=\sigma$, then Eq.~\eqref{eq:KPZ} predicts a 
constant growth velocity $v=\langle\partial_t h\rangle= v_0+\lambda\sigma^2/2$.
Matching the discrete and continuum results than leads to $v_0=1/2$ and $\lambda=-1$.
We verify this behavior by initializing our simulation with a fixed slope, as depicted in the inset of Fig.~\ref{fig:lambda}.
By then measuring the growth velocity $v$ for various slopes $\sigma$, we observe a quadratic dependence,
and confirm a negative value of the nonlinearity $\lambda$, as demonstrated in Fig.~\ref{fig:lambda}.

\subsection{Influence of front profile on composition}

The couplings proportional to $\beta_1$, $\beta_2$ and $c$ encapsulate how height fluctuations alter the selective pressure between the two species, leading to composition field dynamics that depend locally on height variations.

The effect of the first nonlinear coupling parameter $\beta_1$ is most dramatic on domain boundaries between wild-type and mutant species ($\nabla f\neq 0$). 
Like for the KPZ nonlinearity, if the height field has a fixed slope at any time $\nabla h(x,t)=\sigma$, the $\beta_1$ term  induces a steady drift of the domain boundary with a magnitude proportional to $\beta_1\nabla f\cdot \sigma$.
For $\beta_1>0$ the boundary will tend to move down a slope, and vice versa for $\beta_1<0$. 

In order to highlight this effect, we allow for a microscopic parameter $\beta_\mu$ in our numerics that biases the identity of newly added occupied sites based on neighbors on the same level~\cite{Drossel2000,Drossel2003}.
Specifically, the identity of a newly occupied site is determined by randomly copying the identity from either of the neighbors in the layer below or from one of the neighbors in the same layer, the latter weighted by a factor $\beta_\mu$. 
Since on slopes we are more likely to have same-layer neighbors,  this accentuates the coupling between the composition field and the height fluctuations.

\begin{figure}
\includegraphics[scale=.46]{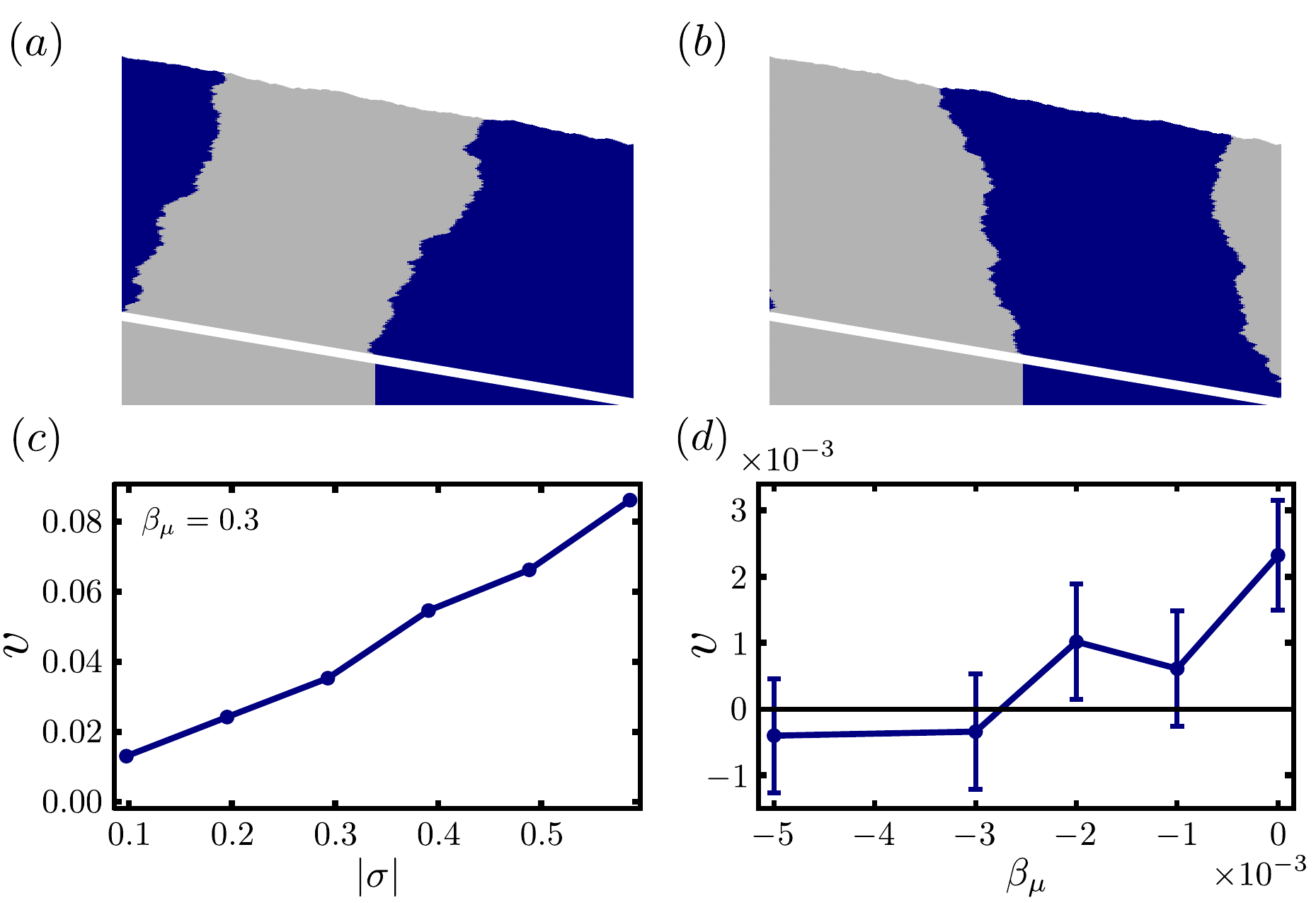}
\caption{Manifestation of the influence of  $\beta_1$  on domain boundary dynamics: 
Simulation of the modified DK algorithm with a nonzero microscopic coupling $(a)$ $\beta_\mu=0.3$, and $(b)$ $\beta_\mu=-0.3$.  The initial condition, demarcated by a white line, has active wild-type sites and inactive mutant sites separated in the middle on a uniform slope $\sigma=-0.5$, and helical boundary conditions
are imposed. $(c)$ Mean domain boundary velocity (lattice spacing/layer) as a function of slope $\sigma$ for fixed microscopic coupling $\beta_\mu=0.3$, obtained from 200 realizations. Error bars smaller than the size of the points. $(d)$ Mean domain boundary velocity for fixed slope $\sigma=-0.5$ as a function of microscopic coupling $\beta_\mu$, with error bars obtained from 250 realizations denoting the standard error on the mean.    Simulations carried out on a lattice with $L=256$ and $\alpha_\mu=\tau_\mu= \mu_\mu=0$.}
\label{fig:beta}
\end{figure}

In Figs.~\ref{fig:beta}$(a)$ and $(b)$, we highlight the influence of $\beta_\mu\neq0$ with two exaggerated examples obtained using sloped initial conditions and a 50/50 split of active wild-type and inactive mutant sites.
For $\beta_\mu>0$, we enhance the drift of the domain boundary down the slope; whereas, by taking a sufficiently negative $\beta_\mu$, we can cause the domain boundary to move up the slope.
We can quantify the macroscopic effect of the model parameter $\beta_\mu$ by measuring the
dependence of the domain boundary velocity on slope as in Fig.~\ref{fig:beta}$(c)$.

It is important to recognize that even in the absence of an explicit microscopic slope to composition coupling 
($\beta_\mu=0$), the modified DK algorithm  still causes any domain boundary to drift with a nonzero speed, {\em i.e.,} the microscopic model has an emergent nonzero macroscopic coupling $\beta_1$ as anticipated
on symmetry grounds in Eq.~(\ref{eq:langevin2}).
We can measure its magnitude, by ramping up small negative $\beta_\mu$ until the mean drift goes away.
We see this in Fig.~\ref{fig:beta}$(d)$ where the mean domain boundary drift was determined from the average of 250 realizations initialized as in Figs.~\ref{fig:beta}$(a)$ and $(b)$.
For $\beta_\mu=0$, the mean domain boundary drift is very small and nonzero.
By making $\beta_\mu$ more negative, we see the mean domain boundary drift passes through zero.
Unfortunately, due to the expense of the simulations our statistics are very poor. 
 Thus, determining the precise crossing point is problematic, but it is reasonable to conclude that its value is in the interval $[-0.005, -0.002]$.

The remaining nonlinear coupling terms in Eq.~(\ref{eq:langevin2}) are proportional to the composition field $f$.
We can interpret their impact by recognizing that they combine to act as an effective control parameter
\begin{equation}
\tau_{\rm eff}=\tau+\beta_2\nabla^2h+\frac{c}{2}(\nabla h)^2.
\end{equation}
The above equation implies that on a sloped surface, the location of the DP transition shifts from that on a flat
surface by an amount that is quadratic in slope. A fit to $\tau_c(\sigma)$ can then be used to estimate
the macroscopic parameter $c$. 
This is reminiscent of depinning of an interface in anisotropic random media in which the quadratic
dependence of critical force on slope can be used to identify a KPZ-type nonlinearity~\cite{TKDprl}.
It is difficult to propose a similar procedure for obtaining the parameter $\beta_2$ as curvature 
changes during the growth process. Variations in composition along a corrugated surface
may provide a means of macroscopically accessing this parameter.

The limit $\beta_1=\beta_2=\beta$ for $c=0$ is interesting, as the terms combine to form an advection, $\beta\nabla (f\nabla h)$, of the field $f({\bf x},t)$ with local velocity proportional to the slope $\nabla h({\bf x},t)$. 
Variations in the integrated composition are 
then not affected by the shape of the surface, although its dynamics can be modified.

\subsection{Dependence of front velocity on composition} \label{sec:slope}

\begin{figure}
\includegraphics[scale=.5]{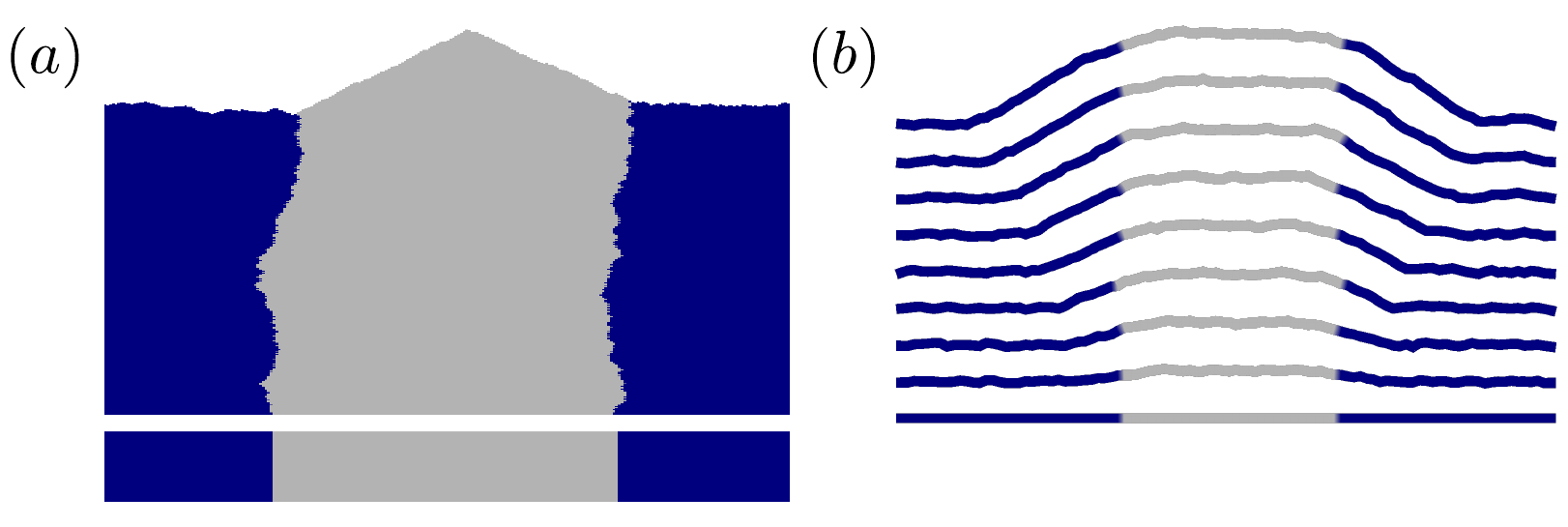}
\caption{Variation of front velocity with composition: $(a)$  For $\lambda<0$, simulation of the modified DK algorithm with a nonzero microscopic coupling that biases the addition of active sites to active regions with probability of $\alpha_\mu=0.2$.  Initial condition, demarcated by white line, is a flat front with half of the region active.  
The biased growth of active sites leads to a prominent triangular bulge in the front.  
Simulations carried out on a lattice of width $L=256$ sites with $\beta_\mu=\tau_\mu= \mu_\mu=0$. 
  $(b)$ For $\lambda>0$, numerical solution of coupled DP/KPZ stochastic partial differential equations, Eqs.~\eqref{eq:langevin1} and \eqref{eq:langevin2} with full demographic noise $\sqrt{f(1-f)}\eta$, using the Euler algorithm with $L=50$.  Pictured are height profiles spaced in time by $135 (\nu/v_0^2)$ colored by composition with gray for $f>1/2$ and blue for $f\le 1/2$. Parameters are $v_0=1.5$, $\nu=0.05$, $\lambda = 1$, $\alpha = 0.5$, $D_h=0.025$, $K=0.05$, $D_f=0.025$, and $\tau=g=\beta_1=\beta_2=0$. }\label{fig:alpha}
\end{figure}

The term proportional to $\alpha$ in Eq.~\eqref{eq:langevin1} represents how differences in expansion rates between active wild-type sites and inactive mutant sites lead to variations in the growth velocity of the film height.
To unravel this effect, we implement in our numerical model the microscopic parameter $\alpha_\mu$, which biases the addition of new occupied sites based on the composition of the local neighborhood.
Specifically, depending on the number of mutant species in the layer below $m_{\rm s}=\{0,1,2\}$, we reject the addition of a new occupied lattice site with probability $m_{\rm s}\alpha_\mu$.
The outcome of an exaggerated implementation with a rather large $\alpha_\mu$, starting with flat initial conditions and a bounded region of active wild-type sites, is presented in Fig.~\ref{fig:alpha}$(a)$.
As time goes on, the preference to add new active sites results in a bulge that quickly assumes a triangular shape, 
while the front above inactive sites remains almost flat. The finite slope of the triangular region originates
in the balance between increased growth speed $\alpha$ with the reduction 
$\lambda\sigma^2/2$ (since $\lambda<0$) due to slope, leading to $\sigma^2=-2\alpha/\lambda$.
Thus the macroscopic parameter $\alpha$ can be estimated from this slope.
For example, in the realization presented in Fig.~\ref{fig:alpha}, we find $\alpha\approx 0.04$.
Note that the above argument relies on a negative value of $\lambda$.  
 To visualize the $\lambda >0$ case, we numerically solved the coupled DP/KPZ stochastic partial differential equations [Eqs.~\eqref{eq:langevin1}  and \eqref{eq:langevin2}] with $\lambda = 1$.
The result is presented in Fig.~\ref{fig:alpha}$(b)$.
Here,  the faster
island remains roughly flat, with triangular dips in the slower domains left behind~\footnote{For
$\lambda=0$, the bulges and dips will have parabolic shapes.}.

Clearly,  differential expansion rates lead to  bulges (or dips) in the front.
The proliferation of such protrusions when both species coexist is expected to increase  the roughness 
of the front, due to greater variation in the height of the jagged domains.
Indeed, this effect has been previously identified in both an agent-based model of the mechanical expansion of reproducing bacteria~\cite{Farrell2017}, as well as in a two-species Eden Model with mutations~\cite{Kuhr2011}, both of which tie the expansion rate to the replication rate.

 \section{Genealogical paths in deterministic growth}

 Additional insight is obtained by examining the deterministic limit.
While much of this section recapitulates known results, by reanalyzing them in this new context we reveal interesting connections between fitness and shape.
Thus, this section can be safely skipped by readers more interested in the scaling analysis.

Our analysis in this section is based on unraveling a hidden connection between surface growth and populations dynamics in the absence of noise.  
To this end, it will prove beneficial to look at a more general model of competition and mutations in biological populations without noise, the Eigen model~\cite{Eigen1971,Eigen1977,Eigen1988, Eigen1989}, whose properties and solution are well documented~\cite{Leuthausser1986,Leuthausser1987,Quasispecies}.
However, to make a connection to the scenario at hand, we here first consider a variant of the model that includes stochasticity
and allows for migration of the populations in space, before returning to the special limit of deterministic growth.

 Consider quite generality a population composed of $N_i({\bf x}, t)$ individuals of $i=1,2,\dots, n$ species
at location ${\bf x}$ at time $t$. The populations change in time according to
\begin{equation}\label{eq:EigenNi}
\partial_t N_i=\nu \nabla^2 N_i+\alpha_i N_i +\sum_j M_{ij}N_j+\sqrt{2DN_i}\eta_i \,, 
\end{equation} 
where  $\eta_i$ are independent, delta-correlated white noises, $\alpha_i$ denotes the fitness (reproductive capacity) of species $i$, and the matrix elements
$M_{ij}$ (with $\sum_i M_{ij}=0$) indicate the rates of mutation from $j$ to $i$.
Equation~(\ref{eq:EigenN}) assumes equal rates of diffusion $\nu$ (migration) and demographic noise $D$ for all species. Note that the overall population size is not fixed and may grow exponentially,
or decay to zero. Indeed, the total population $N({\bf x}, t)=\sum_i N_i({\bf x}, t)$ at  location ${\bf x}$
changes according to
\begin{equation}\label{eq:EigenN}
\partial_t N=\nu \nabla^2 N+\overline{\alpha} N +\sqrt{2DN}\eta \,,
\end{equation}
where $\overline{\alpha}=\sum_i\alpha_i N_i/N$ is the average fitness over the population and $\eta$ is delta-correlated, white noise.

 We can reformulate Eqs.~\eqref{eq:EigenNi} and \eqref{eq:EigenN} in terms of the dynamics of  composition fields, 
as well as an effective height field.
This will allow us to tweeze apart the effects of topography on fitness, in the deterministic limit.  
To this end, we form population fraction fields  $f_i({\bf x}, t)=N_i({\bf x}, t)/N$, akin to our composition field,  
that dynamically couple to what we will designate as a `height field' $h({\bf x}, t)=\frac{2\nu}{\lambda}\ln N$ 
(which is recognizable as the Cole-Hopf transformation).
Now, using Ito's Lemma we find for the variation of the composition fields
\begin{align}\label{eq:Eigenfi}
\partial_t f_i&=\nu \nabla^2 f_i+\lambda\nabla f_i\cdot\nabla h +(\alpha_i- \overline{\alpha})f_i &\\ \nonumber
&+\sum_j M_{ij}f_j+2Df_i(1-f_i)/N+\sqrt{2D/N}\eta^\prime_i,
\end{align}
whose stochasticity can be represented with correlated multiplicative noise $\langle \eta^\prime_i(t)\eta^\prime_j(s)\rangle=(f_i\delta_{ij}-f_{i}f_j)\delta(t-s)$, and 
\begin{equation}\label{eq:EigenLN}
\partial_t h=\frac{2\nu}{\lambda}\overline{\alpha}-\frac{2\nu D}{\lambda N} +\nu \nabla^2 h+\frac{\lambda}{2} (\nabla h)^2+\frac{2\nu}{\lambda}\sqrt{\frac{2D}{N}}\eta.
\end{equation}

The stochastic element in Eq.~\eqref{eq:EigenNi}, proportional to $\sqrt{N_i}$, is due to random
reproduction events (demographic noise). Interestingly, this type of noise goes to zero in Eqs.~\eqref{eq:Eigenfi}
and \eqref{eq:EigenLN} for large population size, in the limit $N\to\infty$.
In this limit, Eq.~(\ref{eq:EigenLN}) becomes the {\it deterministic} KPZ equation~\cite{Kardar}
\begin{align}\label{eq:KPZdet}
\partial_t h=\nu \nabla^2 h+{\lambda\over 2} (\nabla h)^2+{2\nu\over \lambda}\overline{\alpha},
\end{align}
and Eq.~(\ref{eq:Eigenfi}) simplifies to
\begin{align}\label{eq:DPdet}
\partial_t f_i=\nu \nabla^2 f_i+\lambda\nabla f_i\cdot\nabla h +(\alpha_i- \overline{\alpha})f_i +\sum_j M_{ij}f_j;
\end{align}
a noiseless, multi-component version of 
Eq.~(\ref{eq:langevin2}) with $K=\nu$, $\beta=\lambda$ and $\beta_2=c=0$.
Indeed for two species ($n=2$) with $f_1=f$ and $f_2=1-f$, $\overline{\alpha}=\alpha_1f+\alpha_2(1-f)$
and mutations only from 1 to 2 at rate $M_{11}=-\mu$, we recover Eq.~(\ref{eq:langevin2})
with $\tau=\alpha_1-\alpha_2-\mu$ and $g/2=\alpha_1-\alpha_2$; also
$\alpha=2\nu(\alpha_1-\alpha_2)/\lambda$ in Eq.~(\ref{eq:langevin1}).
 Thus, within the  Eigen model we find a special limit of the competition dynamics we have been studying.

\begin{figure}
\includegraphics[scale=.61]{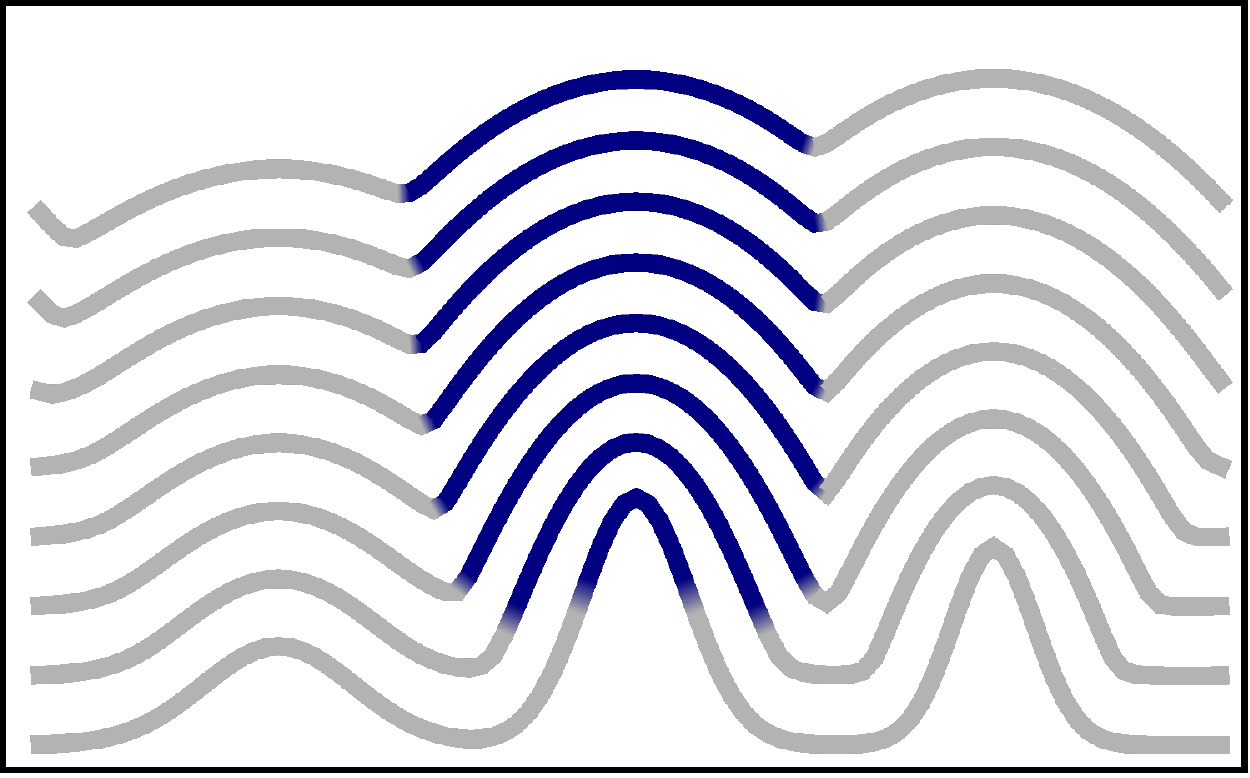}
\caption{Illustration of deterministic coarsening dynamics: The profile moves up by taking the envelope of a collection of parabolic mounds emitted from each location on the surface.  The bacterial composition field (pictured in blue and gray) follows the surface coarsening dynamics deterministically, with the order parameter value at the local maximum dominating the neighborhood.
 The blue bacteria despite having a slightly lower reproductive fitness ($\alpha_{\rm blue} <\alpha_{\rm gray}$) are able to take over the local neighborhood.  Illustration generated from a numerical solution of Eqs.~\eqref{eq:KPZdet} and \eqref{eq:DPdet} using the Euler algorithm with $L=50$, $v_0=0.5$, $\nu=K=0.1$, $\lambda=1$, $\alpha=0.06$, $\tau=g/2=0.3$, $\beta=1$.}\label{fig:coleHopf}
\end{figure}

The Cole-Hopf transformation maps the noiseless (deterministic) but nonlinear Eqs.~\eqref{eq:KPZdet} and \eqref{eq:DPdet}
for $h$ and $f_i$ into a set of linear equations that are easily solved~\cite{Kardar1999}.
Given an initial population of composition $f_i({\bf x}, 0)$ distributed on a landscape
characterized by a profile $h({\bf x}, 0)$ as in Fig.~\ref{fig:coleHopf}, we can construct 
`populations' $N_i({\bf x}, 0)=f_i({\bf x}, 0)\exp\left[\frac{\lambda}{2\nu}h({\bf x}, 0)\right]$.
The deterministic version of Eq.~(\ref{eq:EigenNi}) can now be easily integrated to
\begin{align}\label{eq:EigenNit}
N_i({\bf x}, t)\!=\!\!\int\! \!\frac{d^d{\bf x}'}{(4\pi\nu t)^{d/2}}& \exp\left({-\frac{|{\bf x}-{\bf x}'|^2}{4\nu t}}\right)
\left[e^{Gt}\right]_{ij}&\nonumber\\
\times &\exp\left[\frac{\lambda}{2\nu}h({\bf x}',0)\right] f_j({\bf x'},0),&\,
\end{align}
where the matrix $G_{ij}=M_{ij}+\alpha_i\delta_{ij}$ combines the effects of mutation and fitness.

 This mapping through Eq.~\eqref{eq:EigenNit} illustrates the importance of inhomogeneous initial conditions,
\emph{e.g.}, uniformly inoculating a non-flat surface with height profile $h({\bf x},0)$, 
or a flat surface with a non-uniform  population number $N({\bf x},0)$.
In the simplest case of non-competitive growth of two species ($\alpha_1=\alpha_2$) with no mutations, \emph{i.e.}, for
$G=0$, the form of the front at long times (alternatively, for $\nu\to 0$) can be extracted from Eq.~(\ref{eq:EigenNit}) using the saddle point 
approximation, as
\begin{equation}\label{eq:saddle}
h({\bf x},t)={2\nu\over \lambda}\ln N\approx \max_{{\bf x}'}\left[h({\bf x}',0)-\frac{|{\bf x}-{\bf x}'|^2}{2\lambda t}\right]\,,
\end{equation}
 In this limit, the surface profile (for $\lambda>0$)
is  a collection of parabolic mounds, each centered on a local peak, say at ${\bf x_0'}$
of the initial profile.
Equation~\eqref{eq:EigenNit} then implies $f_i({\bf x},t)=f_i({\bf x_0'},0)$, 
\emph{i.e.}, that the entire ``mound" is populated by decedents 
of the quasi-species member residing at the initial local peak ${\bf x_0'}$. 
The advantage conferred by the initial profile in this model is
equivalent to an increased population size  (on a flat surface) proportional to $\exp[\frac{\lambda}{2\nu}h({\bf x},0)]$.
As a result, quasi-species at the height of a local maximum for $\lambda>0$, or a minimum for $\lambda<0$, slowly take over neighboring regions as the coarsening progressively spreads out.

 The emergence of parabolic mounds, as well as the benefits of initial location to survival, are illustrated in
Fig.~\ref{fig:coleHopf}. Here, the species starting at the highest peak is in fact the less fit ($\alpha_{\rm blue} <\alpha_{\rm gray}$), and would disappear with a flat starting profile. However, starting from a high peak, it actually expands at the
expense of the fitter species at lower height, establishing a niche within the growing parabola.

More generally, the combined diffusion/growth/mutation kernel in Eq.~(\ref{eq:EigenNit}) can also be 
cast as an integral over paths, 
\begin{align}\label{pathI}
&\frac{1}{(4\pi\nu t)^{d/2}} \exp\left({-\frac{|{\bf x}-{\bf x}'|^2}{4\nu t}}\right)
\left[e^{Gt}\right]_{ij}\Longrightarrow\\
&\int^{({\bf x},t)}_{({\bf x}',0)}\!{\mathcal D}[{\bf x}(t')]\!\exp\left[-\frac{1}{4\nu}\int_0^tdt'\!\left(\frac{d{\bf x}(t')}{dt'}\right)^2\right] \nonumber  \\
&\qquad\times \prod_{m=1}^n e^{\alpha_{i(m)}(t_m-t_{m-1})}\,\prod_{m=0}^n M_{i(m),i(m+1)}\,.\nonumber
\end{align}
The first (diffusion) component of the kernel is dominated by paths that proceed along a
straight line from ${\bf x}'$ to ${\bf x}$ in time $t$, with `velocity' ${\bf v}=d{\bf x}/dt=({\bf x}-{\bf x}')/t$.
For $G=0$, as in Fig.~\ref{fig:coleHopf}, such straight paths join points on a parabolic segment to
the corresponding local maximum,  indicating the most likely genealogical history 
for the species at the front. 
In the presence of mutations and fitness variations, genealogical tracks are composed of segments
$i=1,2\cdots,n$. In the $m^{th}$ segment, from $t_{m-1}$ to $t_m$, ancestry is of quasi-species $i(m)$,
with $i(0)=j$ and $i(n)=i$.  
The first factor in the third line of Eq.~\eqref{pathI} represents the increase in population size according
to the fitness within each segment, the second factor accounts for the mutations probabilities between segments.

The path integral formulation can be used to explain the jagged profiles observed in Fig.~\ref{fig:alpha}. 
Consider two species starting from a flat profile as in this figure, with initial separation point marked as $x=0$,
 with different fitness values $\alpha_1>\alpha_2$.
The optimal genealogical track terminating at $(x,t)$ in quasi-species 2 may actually correspond to ancestry 
starting with quasi-species 1 at $x=0$, proceeding vertically along $x=0$ for time $t-\tau$,  followed by a portion 
that proceeds with velocity $v=x/\tau$ to the target point in the remaining time $\tau$ after transforming to 
quasi-species 2.
The vertical segment gains weight  $\alpha_1(t-\tau)$ in Eq.~\eqref{pathI},
the sloped segment has cost $\alpha_2\tau-{x^2\over 4\nu\tau}$. 
The resulting height profile is obtained by optimizing the sum of the two contributions as
\begin{eqnarray}\label{eq:PathSlope}
h({x},t)&=&{2\nu\over\lambda}\max_{\tau}\left[\alpha_1(t-\tau)+\alpha_2\tau-\frac{x^2}{4\nu\tau}\right]\nonumber\\
&=& v_1 t-\min_{\tau}\left[\frac{x^2}{2\lambda\tau}+\alpha\right] \,,
\end{eqnarray}
where we have noted $\alpha=2\nu(\alpha_1-\alpha_2)/\lambda$, and set $v_1=2\nu\alpha_1/\lambda$.
There is an optimum at $\tau^*={|x|\over\sqrt{4\nu(\alpha_1-\alpha_2)}}$, which upon substitution
in the above equation leads to a slope of  $\sigma^2=h^2/x^2=|2\alpha/\lambda|$,
consistent with the result in Sec.~\ref{sec:slope}.

\section{ Special limits and symmetries}\label{sec:maps}

The remainder of this paper focuses on  universal scale-invariant properties of  
Eqs.~\eqref{eq:langevin1} and \eqref{eq:langevin2}.
However, for certain choices of parameters these equations posses symmetries,
providing additional constraints and insights on their general properties. 
 While this information is familiar knowledge from earlier studies on interface roughening and directed percolation, it is
nevertheless  valuable to present the results in this context.

\subsection{`Galilean invariance' and scaling of roughness}
There is no characteristic (microscopic) scale limiting the extent of roughness, which thus assumes
scale-invariant characteristics. For a single-valued height function, this leads to self-affine scaling,
and dynamic fluctuations characterized by the scaling form~\cite{Kardar}
\begin{equation}\label{eq:KPZScale}
\langle [h({\bf x},t)-h({\bf x}^\prime,t^\prime)]^2\rangle = |{\bf x}-{\bf x}^\prime|^{2\chi}g\left(\frac{|t-t^\prime|}{|{\bf x}-{\bf x}^\prime|^{z_h}}\right),
\end{equation}
where $\chi$ is the roughness exponent (later also designated as $y_h$),
and $z_h$ is a dynamic scaling exponent, measuring the relative scalings in space and time.
For the one dimensional KPZ equation, $\chi=1/2$ and $z_h=3/2$ are exact results~\cite{Kardar1986},
while in two dimensions, the estimates $\chi_h\approx0.4$ and $z_h\approx1.6$ come from
numerical simulations~\cite{H-HK15}.

In the absence of the coupling to the composition field, the KPZ equation's scaling exponents are not independent, due to a hidden symmetry, which is a remnant of the rotational symmetry of the gradient expansion leading to the KPZ equation.
Defining a velocity field related to the slope by ${\bf v}=-\lambda{\bf \nabla}h$~\cite{Kardar1986,Frey1994},
maps Eq.~(\ref{eq:langevin1}) (for $\alpha=0$) to the Burgers equation~\cite{BurgersBook}
\begin{equation}\label{eq:KPZsym}
\frac{D {\bf v}}{Dt}\equiv\partial_t{\bf v}+{\bf v}\cdot{\bf \nabla v}=\nu\nabla^2{\bf v}-\lambda  \sqrt{2D_h}{\bf \nabla}\xi,
\end{equation}
which describes a randomly stirred fluid.
The significance of this mapping is that the fluid flow equation is (statistically) invariant under a 
Galilean transformation to a moving frame (${\bf v'}({\bf x'},t')={\bf v}({\bf x},t)+{\bf u}$ with
${\bf x'}={\bf x}+{\bf u}t$ and $t'=t$). In terms of the KPZ equation, this transformation corresponds
to a tilt
\begin{equation}\label{eq:KPZsym}
h^\prime=h+{\bf \epsilon\cdot x},\quad {\bf x^\prime}={\bf x}-\lambda {\bf \epsilon} t,\quad t^\prime=t,
\end{equation}
with ${\bf u}=-\lambda {\bf \epsilon}$.
The significance of the Galilean symmetry is that it constrains the scaling of ${\bf v}$ and hence ${\bf \nabla}h$
in terms of those of space and time (as ${\bf x}/t$), leading in Eq.~(\ref{eq:KPZScale}) to the exponent
identity $\chi=2-z_h$. 
Formally, since $\lambda$ appears both in the symmetry mapping of Eq.~\eqref{eq:KPZsym} and as the nonlinearity in the KPZ equation, any renormalization scheme that maintains the symmetry cannot renormalize $\lambda$ (at any order).

An infinitesimal transformation as in Eq.~(\ref{eq:KPZsym}), leads to $\nabla_{\bf x}=\nabla_{\bf x'}$
and $\partial_t=\partial_t'-\lambda{\bf \epsilon\cdot \nabla_{\bf x'}}$ with $\nabla_{\bf x}h=\nabla_{\bf x'}h'-{\bf \epsilon}$. It is easy to check that these transformations also leave Eq.~(\ref{eq:langevin2}) invariant
if $\beta_1=\lambda$ and $c=0$ (for all values of $K$ and $\beta_2$). 
With $\beta_1=\lambda$, both the composition  and front--slope fluctuations are  transported
along the front with the same speed and maintained in phase. The dynamics of composition
fluctuations should thus mirror those of the front, when there is rotational symmetry.

\subsection{Scaling in directed percolation}

The transition to an absorbing state at a critical value of $\tau_c$ is accompanied by 
a number of singularities. These include divergences of the correlation lengths in space $\xi_\perp$ -- perpendicular to the direction of growth -- and time $\xi_\parallel$ -- parallel to the growth -- as well as, the vanishing of density of particles in the active phase $\rho$, and the probability 
of a single active seed percolating through space $P_{\rm perc}$. In the limit of $\delta\tau=\tau-\tau_c\to0$, the scaling of these quantities are characterized by critical exponents as~\cite{Henkel}
\begin{equation}\label{eq:DPScale}
\xi_\perp\sim\delta\tau^{-\nu_\perp},\ \xi_\parallel\sim\delta\tau^{-\nu_\parallel},\ \rho\sim\delta\tau^{\beta},\  P_{\rm perc}\sim\delta\tau^{\beta^\prime}.
\end{equation}
A dynamic critical exponent, relating the scalings in the two directions, is defined by
$z_f=\nu_\parallel/\nu_\perp$.
The values of these exponents have been obtained previously from a renormalization group  analysis near the critical dimension $d_c=4$, perturbatively in $\epsilon=4-d$, and are listed in Table~\ref{table:exp}~\cite{Tauber2005,Tauber}.
In addition for comparison, we have listed values at $d=1$, which were calculated using a series expansion \cite{Jensen1996}.

The DP equation also satisfies a form of time-reversal symmetry known as rapidity-reversal~ \cite{Henkel}.
It is manifest only at the level of the dynamic action, which we will discuss in the following sections.
Briefly, if we introduce the response field ${\tilde f}(x,t)$ conjugate to the order parameter $f(x,t)$, then rapidity-reversal is a symmetry under the exchange
${\tilde f}(x,t)\longleftrightarrow -f(x,-t)$
of the dynamic action.
This symmetry has two important consequences.
The first is that it imposes the exponent identity $\beta=\beta^\prime$: the disappearance of a cluster of active wild-type sites is the time reverse of the percolation of a cluster of active wild-type sites from one active seed~\cite{Henkel}.
Second, on a more practical level, this symmetry forces $g$ and $D_f$ to renormalize together, implying that there is actually only one relevant parameter $u\propto gD_f$.
Since rapidity-reversal is a relative of time-reversal symmetry -- a symmetry not shared by the KPZ equation --, we do not expect this symmetry to remain for our equations, as manifested in $g$ and $D_f$ renormalizing differently with unequal $\beta$ and $\beta^\prime$ exponents.

\section{Critical exponents and renormalization}\label{sec:RG}

We now turn to  the analysis of the universal scaling behaviors at the active-to-absorbing phase transition in the presence of surface fluctuations.  

\subsection{Setup}

Our renormalization group (RG) analysis is based on the  Martin-Siggia-Rose-Janssen-de Dominici (MSRJD) dynamic functional \cite{Tauber}, which plays the role of the Hamiltonian in a nonequilibrium renormalization scheme.
To this end, we  determine the probability distribution of the space-time trajectories of $f$ and $h$ by noise-averaging $\delta$-functions that enforce the equations of motion [Eqs.~\eqref{eq:langevin1} and \eqref{eq:langevin2}]
\begin{equation}
{\mathcal P}[f,h]=\left\langle \delta(f-f[\eta])\delta(h-h[\xi])\right\rangle_{\xi,\eta}\, ,
\end{equation}
where the average is over Gaussian noise
\begin{equation}
{\mathcal P}[\xi,\eta]\propto \exp\left\{-\int d^d{\bf x}dt\left[ \frac{\xi({\bf x},t)^2}{2}+\frac{\eta({\bf x},t)^2}{2}\right]\right\}.
\end{equation}
The $\delta$-functions are then replaced with their generalized Fourier representations by introducing conjugate (response) fields ${\tilde h}$ and ${\tilde f}$, allowing us to cast the probability distribution into the form
\begin{equation}
{\mathcal P}[f,h]\propto\int{\mathcal D}[\tilde f]{\mathcal D}[{\tilde h}]\ \exp\left\{-{{\mathcal J}[{\tilde f},f,\tilde{h},h]}\right\}.
\end{equation}
The MSRJD function naturally splits as ${\mathcal J}={\mathcal J}_0+{\mathcal J}_{\rm int}$, into a  solvable Gaussian contribution
\begin{equation}\label{eq:Gaus}
\begin{split}
{\mathcal J}_0=\int d^{d}{\bf x} dt\ \Big[& {\tilde f}\left(\partial_t f -K\nabla^2f -\tau f \right)\\
&+{\tilde h}(\partial_th-\nu\nabla^2h-\alpha f)-D_h{\tilde h}^2\Big],
\end{split}
\end{equation}
and a piece containing all higher-order interactions
\begin{equation}\label{eq:nonLinear}
\begin{split}
{\mathcal J}_{\rm int}=\int d^{d}{\bf x} dt\ \bigg[ &\frac{g}{2}{\tilde f}f^2-D_ff{\tilde f}^2 -\beta_1{\tilde f}\nabla f\cdot\nabla h \\
&-\beta_2{\tilde f}f\nabla^2h-\frac{c}{2}{\tilde f}f(\nabla h)^2 -\frac{\lambda}{2}{\tilde h}(\nabla h)^2\bigg].
\end{split}
\end{equation}

The first step in our scaling analysis is to obtain the naive scaling dimensions of all the parameters.
To this end, we rescale space, time, height and order parameter according to 
$x\to b x$, $t\to b^z t$, $f\to b^{y_f}f$, $h\to b^{y_h}h$, ${\tilde f}\to b^{{\tilde y}_f}{\tilde f}$, and ${\tilde h}\to b^{{\tilde y}_h}{\tilde h}$, and then require the invariance of the Gaussian contribution ${\mathcal J}_0$.  This procedure leads to the mean-field values $z=2$,  $y_f={\tilde y}_f=-d/2$, $y_h=2-d/2$ and ${\tilde y}_h=-2-d/2$.
With these scalings, all nonlinearities acquire a scaling dimension $2-d/2$, with a critical dimension $d_c=4$.
However, if $\alpha\neq0$ the noise in the KPZ equation $D_h$ scales as 
$d+z+2{\tilde y}_h=-2$ and is irrelevant in any dimension, and will be ignored when this is the case.
Furthermore, the nonlinear coupling proportional to $c$, while possibly relevant, is 
higher order in the sense of not renormalizing to one-loop order, and we shall also ignore it.

To carry out the RG calculation, we employ Wilson's momentum shell approach~\cite{Tauber2012,Kardar}.
This is accomplished by first integrating out short wavelength (high momentum) modes $\Lambda/b < |{\bf q}|<\Lambda$, where $\Lambda$ is the wave vector cutoff and $b>1$ is a scaling factor.
We then rescale space, time, height and order parameter as before.
This requires treating the nonlinearities perturbatively by expanding $e^{-{\mathcal J}_{\rm int}}$ in powers of the couplings.
Finally, taking $b=e^l$ with $l\ll1$, leads to the differential RG equations for the model parameters.
To be consistent with the perturbation expansion, we have to focus on dimensions near the critical dimension, so we set $d=4-\epsilon$ with $\epsilon\ll1$.
This technique is standard~\cite{Tauber2012,Kardar}, so we do not go into more detail here, simply collecting the results in the  Appendix~\ref{sec:RGeqs}.

Before turning to the analysis of the fully coupled dynamics, we will first examine two limiting cases with unidirectional coupling between the fields that allow us to highlight the subtle phenomenology involved in this nonequilibrium critical transition.

\subsection{ Profile independent competition, rough growth front,  with feedback }\label{sec:isoNoFeed}

We first consider profile independent competition, where surface dynamics does not influence the DP order parameter:
 \begin{align}\label{eq:isotropic1}
\partial_t h &= \nu \nabla^2 h +\frac{\lambda}{2}(\nabla h)^2+\alpha f \,,\\ \label{eq:istropic2}
\partial_t f &= K\nabla^2 f+\tau f-\frac{g}{2}f^2 +\sqrt{2D_f f}\eta\,. 
\end{align}
In this case, the absorbing phase transition is unaltered by the height field, though the presence of order-parameter fluctuations is expected to alter the scaling properties of the height field, with the composition field acting as a source of correlated noise~\cite{Medina1989}.
The general RG equations presented in the Appendix~\ref{sec:RGeqs} then reduce to the following
\begin{align}
\frac{d\nu}{dl}&=(z-2)\nu\,,\label{eq:RGnu}\\
\frac{d\lambda}{dl}&=\left(z-2+\chi\right)\lambda\,,\label{eq:RGlambda}\\
\frac{d\alpha}{dl}&=\left(z-\chi+y_f\right)\alpha+\frac{\lambda \alpha^2}{2\nu}\frac{A_{4}D_f}{K(K+\nu)}\,.\label{eq:RGalpha}
\end{align}

Above the DP critical dimension, $d>4$, simple dimensional analysis of the DP equation yields $y_f=-d/2$ and $z=2$. 
The choice of $\chi=(2-d/2)<0$ corresponds to the Gaussian fixed point for the field $h$,
at which small $\lambda$ is irrelevant and flows to zero. 
There is also, however, an unstable fixed point with $\chi=0$ for 
$(\lambda\alpha)^*\propto(d-4)>0$.
Larger values of $(\lambda\alpha)$ lead to a rough phase that is not perturbatively accessible. 
While a similar transition exists for white noise in the KPZ equation (for $d>2$), the difference
is that here the transition is controlled by the product  $(\lambda\alpha)>0$. 

We may anticipate that after the unstable fixed point collides with the Gaussian fixed point at $d=4$,
a stable fixed point  emerges for $d<4$. 
However, the scaling exponents at the DP fixed point (listed in 
Table~\ref{table:exp}~\cite{Tauber2005,Tauber})
are $z_f=2-\epsilon/12$ and $y_f=-2+7\epsilon/12$. 
This value of $z$ in Eq.~(\ref{eq:RGnu}) pushes $\nu$ towards zero leading
to a divergence of the correction term in Eq.~(\ref{eq:RGalpha}).

The complexity of this situation is best understood by first considering the limit $\lambda=0$. 
In this case, a perturbation of the field $f$ at length scale $\ell$ will persist for a time
$\tau_f(\ell)\sim A_f\ell^{z_f}$. The amplitude $A_f$ is modified from the Gaussian value of $1/K$
by renormalization of non-linear couplings to $A_f\sim \frac{1}{K}\left(\frac{gD_f}{K^2}\right)^{2-z_f}$.
With $\lambda=0$, the response of the height field to the perturbation ($\alpha f$) is linear, and
characterized by a time-scale $\tau_h\sim \ell^2/\nu$.
The coupled equations are thus characterized by two time scales that are matched at a non-universal
length scale $\ell_c\sim (\nu A_f)^{\frac{1}{2-z_f}}$.
For $\ell\ll\ell_c$ height perturbations are smoothed out by the Laplacian term in the front equation,
while distortions at scales $\ell>\ell_c$ are created and removed rapidly by the field $f$ of critical DP.
In each case, we can associate a roughness for the profile (by setting $d\alpha/dl=0$) as
$\chi=z+y_f$, leading to $\chi=7\epsilon/12$ for $\ell<\ell_c$ ($z=2$), and
$\chi=\epsilon/2$ for $\ell>\ell_c$ ($z_f=2-\epsilon/12$).
This possibility where dynamic scaling exponents between fields can be different is known as weak dynamic scaling -- to be distinguished from strong dynamic scaling when the exponents are the same.
Such an effect is not unusual when coupling multiple fields together~\cite{Folk2006} and has been observed in Models C and D~\cite{Akkineni2004}, kinematic waves~\cite{Das2001}, active-to-absorbing phase transitions in fluctuating environments~\cite{Sarkar2012,Sarkar2014}, as well as for conserved KPZ equation coupled to critical fields~\cite{Banerjee2018}.

For $\lambda\neq0$, the choice of $z=2$ and $\chi=0$ indicates that a small $\alpha$
is relevant, and the  second order term in Eq.~(\ref{eq:RGalpha}) has to be considered.
Now for $(\lambda\alpha)<0$ a stable fixed point appears with the above exponents. 
For $(\lambda\alpha)>0$ the coupling runs to large values, making it impossible to assign effective
exponents to the regime $\ell<\ell_c$ by the current analysis. 

\subsection{ Profile dependent competition, rough growth front, without feedback}

Whereas the previous section  considered the field $f$ evolving without influence
from the front profile, here we consider that the dynamics of profile $h$ gets no feedback from
the composition field while affecting its variations according to 
\begin{align}
\partial_t h &= \nu \nabla^2 h +\frac{\lambda}{2} (\nabla h)^2 + \sqrt{2D_h}\xi \,, \label{eq:KPZ2} \\ \nonumber
\partial_t f &= K\nabla^2 f+\tau f -\frac{g}{2}f^2 +\sqrt{2D_f f}\eta\\ 
&\qquad+\beta_1\nabla f\cdot \nabla h +\beta_2 f \nabla^2 h +\frac{c}{2}f(\nabla h)^2.\label{eq:DP2}
\end{align}
We have included the factor proportional to $c$. Even though we do not consider it in the full RG treatment; the following conclusions nevertheless apply. 

As far as the  perturbative RG analysis is concerned, the behavior of the above coupled equations are either trivial or inaccessible. This is because in dimensions $d>2$, the nonlinearity in Eq.~(\ref{eq:KPZ2}) is either irrelevant
leading to roughness exponent $\chi=(2-d)/2<0$, or relevant but flowing to uncontrolled large values.
In the latter case perturbative RG in not useful, leaving numerical simulations as a promising approach~\cite{Kuhr2011}.
In the former case, negative values of $\chi$ imply that the three non-linear couplings $(\beta_1,\beta_2,c)$,
involving factors of $h$ or $h^2$, in Eq.~(\ref{eq:DP2}) are irrelevant.
With all non-linearities eventually disappearing at large scale, the behavior then mimics uncoupled roughness
and DP.

\subsection{Full renormalization group analysis}

Having discussed how unidirectional coupling affects the scaling properties of the DP transition, 
or the surface roughness, we now consider an RG analysis of the fully coupled rough DP equations in $d=4-\epsilon$.

The RG equations turn out to only  depend on the following reduced set of dimensionless effective parameters, whose analysis is substantially simpler:
\begin{align}\label{eq:effPara}\nonumber
\theta=&\frac{K}{\nu},\qquad\ \   \bar\tau=\frac{\tau\Lambda^2}{K},\\ 
u=\frac{gD_fA_4}{K^2},\ &\zeta =\frac{\lambda\alpha D_fA_4}{K\nu^2},\ \delta_j=\frac{\beta_j\alpha D_f A_4}{K^2\nu},
\end{align}
with $A_4=S_4/(2\pi)^4$ where $S_4$ is the surface area of a 4-dimensional sphere and is included for convenience.
In the first row, $\theta$ measures the relative relaxation times of the DP equation to the KPZ equation, and $\bar\tau$ is 
the dimensionless control parameter for the DP transition.
The remaining parameters are the dimensionless combinations of nonlinear couplings.  
Importantly, by dividing out the $K$ and $\nu$ dependencies, they capture solely their effects on spatial variations, 
leaving the relative temporal dynamics encoded in $\theta$.

The fixed point structure further is controlled by the value of the time-scale ratio $\theta=K/\nu$.
Since $\theta$ is not a nonlinear coupling, it need not be perturbatively small.
Thus, there are three possible phenomenologically distinct fixed points 
corresponding to $\theta=0$, $\theta\to\infty$, or with a nonzero finite value at the fixed point.
We address each possibility in turn.

\subsubsection{$\theta\to\infty$  sector}

This limit represents the situation where $K\gg\nu$, which signifies that the DP dynamics evolve on a much faster time scale than the KPZ equation.  
Put another way, from the perspective of  surface fluctuations, the DP equation relaxes instantaneously to its probabilistic stationary state conditioned on the surface configuration.

In this limit, the RG equations for the nonlinear couplings reduce to
\begin{align}
\frac{du}{dl}&=u(\epsilon-3u/2)\\
\frac{d\zeta}{dl}&=\zeta(\epsilon-5u/8)\\
\frac{d\delta_1}{dl}&=\delta_1(\epsilon-3u/4)\\
\frac{d\delta_2}{dl}&=\delta_2(\epsilon-\delta_2)+u\delta_1/4\\
\frac{d\bar\tau}{dl}&=\bar\tau\left(2-u/8\right)-u/2.
\end{align}
There are two fixed point solutions to these equations, where all derivatives are zero.
The first is the Gaussian fixed point: $u^*=\zeta^*=\delta_1^*=\delta_2^*=0$.
This corresponds to the uncoupled `mean-field' behavior, which is stable above the critical dimension $d_c=4$, but unstable below.
The second solution corresponds to the standard DP fixed point: 
$u^*=2\epsilon/3$ and $\zeta^*=\delta_1^*=\delta_2^*=0$ (cf. Table~\ref{table:exp}).
Here, the surface fluctuations are driven by the DP fluctuations, but not vice versa.
As in Sec.~\ref{sec:isoNoFeed}, we can explore different scalings and responses of height
fluctuations at short and long scales; however,  the 
couplings ($\beta_1$ and $\beta_2$) are relevant, rendering the standard DP fixed point unstable.

\subsubsection{$\theta$ zero sector}

When $\theta=0$,  we have the reverse situation where $K\ll\nu$, and the surface  dynamics evolve much faster than the
DP composition field.
In this limit, the RG equations for the dimensionless nonlinear couplings become
\begin{align}
\frac{du}{dl}&=u(\epsilon-3u/2+5\delta_1/2-\delta_2)+2(\delta_1-\delta_2)\delta_2\\
\frac{d\zeta}{dl}&=\zeta(\epsilon+\zeta/2-5u/8+\delta_1-5\delta_2/4)\\
\frac{d\delta_1}{dl}&=\delta_1(\epsilon+\zeta/2-3u/4+\delta_1-3\delta_2/2)-\zeta\delta_2/4\\
\frac{d\delta_2}{dl}&=\delta_2(\epsilon+\zeta/2-u+7\delta_1/4-2\delta_2)+\delta_1(u-\zeta-\delta_1)/4\\ \label{eq:tau}
\frac{d\bar\tau}{dl}&=\bar\tau(2+u/8-\delta_1/2+\delta_2/4)-u/2+(\delta_1-\delta_2)/2\,.
\end{align}
Using a symbolic Mathematics package, we have found 12 fixed points for these equations, including the Gaussian fixed point and 
the uncoupled DP fixed point (cf. Sec.~\ref{sec:isoNoFeed}).

Of all the fixed points, a linear stability analysis reveals only one to be both stable and in the physical parameter regime ($u > 0$)~\footnote{There is one fixed point for which all nonlinear couplings are stable and $\theta$ is marginally stable ($d\theta/dl|_{\theta=0}=0$), but flows $\theta\to\infty$ for any finite value of $\theta$: $u=0$, $\zeta=2\epsilon$, $\delta_1=-2\epsilon$, and $\delta_2=0$.  To verify the stability of this fixed point, higher order terms need to be considered.}, with
\begin{equation}
u^*=8\epsilon/3,\ \zeta^*=0,\ \delta_1^*=\delta_2^*=-2\epsilon,\ \bar\tau^*=2\epsilon/3.
\end{equation}
The accompanying scaling exponents are obtained by demanding scale invariance of the remaining parameters, whose RG equations at this fixed point are
\begin{align}\label{eq:nu}
\frac{d\nu}{dl}&=(z-2)\nu\\
\frac{dK}{dl}&=(z-2-\epsilon/6)K\\ \label{eq:D}
\frac{dD_f}{dl}&=(z-y_f-4-\epsilon/3)D_f\\  \label{eq:alpha}
\frac{d\alpha}{dl}&=(z+y_f+{\tilde y}_h+4-\epsilon)\alpha,
\end{align}
combined with the algebraic constraints on the field exponents
\begin{align}\label{eq:alg1}
&y_h+{\tilde y}_h+4-\epsilon=0\\ \label{eq:alg2}
&y_f+{\tilde y}_f+4-5\epsilon/3=0.
\end{align}

As discussed in Sec.~\ref{sec:isoNoFeed}, we have a weak dynamic scaling fixed point with two possible choices for the dynamic scaling exponent, $z=2$ or $2+\epsilon/6$, both of which are consistent with the requirement that $\theta=K/\nu\to 0$ is a stable fixed point.
If we choose $z=2$, then $\nu$ is scale invariant ($d\nu/dl=0$), and $K$ decays as $K\sim e^{-\epsilon l/6}$.
However, this behavior is at odds with the underlying premise of RG that we isolate the scale-invariant behavior that emerges on the longest length and time scales. 
With this choice of $z$, the RG procedure ends up holding the dynamics on the time scale of $\sim1/\nu$ fixed.
But in so doing, the RG procedure commensurately keeps slowing down the DP dynamics controlled by the time-scale $\sim1/K$.
This is a consequence of the the fact there is actually a slower time scale controlled by $K$.
Thus, this choice of $z$ does not focus us  on the slowest time scale,
where the appropriate choice is $z=2+\epsilon/6$.
We should note that as observed in Sec.~\ref{sec:isoNoFeed}, this choice does not rule out another dynamic scaling exponent that controls the response to perturbations on shorter wavelengths.

With this caveat in mind, the rest of the scaling exponents now follow by making $D_f$ and $\alpha$ [Eqs.~\eqref{eq:D} and \eqref{eq:alpha}] scale invariant, coupled with the algebraic constraints in Eqs.~\eqref{eq:alg1} and \eqref{eq:alg2}, as well as extracting the growth rate of $\bar\tau$ from Eq.~\eqref{eq:tau}; leading to
\begin{align}
&y_f=-2-\epsilon/6,\quad {\tilde y}_f=-2+11\epsilon/6 ,\\
&y_h=-\epsilon/6,\quad {\tilde y}_h=-4+7\epsilon/6,\\
&y_\tau=2+5\epsilon/6.
\end{align}
However, the exponents most often measured are those defined in Eqs.~\eqref{eq:KPZScale} and \eqref{eq:DPScale}.
They are obtained through the scaling relationships: $\nu_\perp=1/y_\tau$, $\beta =- y_f/y_\tau$, $\beta^\prime =- {\tilde y}_f/y_\tau$, and $\chi=y_h$~\cite{Henkel}.

We have collected our results in Table~\ref{table:exp}.
Included there for comparison are values obtained from a field-theoretic analysis of the uncoupled DP phase transtion in $d=4-\epsilon$~\cite{Henkel}, as well as exponents measured in simulations of a two-species Eden Model in $d=1$~\cite{Kuhr2011}, which is expected to be in the DP/KPZ universality class.
\begin{table}
\begin{ruledtabular}
\begin{tabular}{l|c|c|c|c|c|c}
Exponent & $z_f$ & $\nu_\perp$ & $\beta$ &$\beta^\prime$ & $z_h$ & $\chi$  \\
\hline
DP/KPZ ($d=4-\epsilon$) & $2+\frac{\epsilon}{6}$ & $\frac{1}{2}-\frac{5\epsilon}{24}$ & $1-\frac{\epsilon}{3}$ & $1-\frac{4\epsilon}{3}$& $2$ & $-\epsilon/6$   \\
&&&&&&\\
DP ($d=4-\epsilon$) & $2-\frac{\epsilon}{12}$ & $\frac{1}{2}+\frac{\epsilon}{16}$ & $1-\frac{\epsilon}{6}$ &$1-\frac{\epsilon}{6}$  &  &   \\
DP ($d=1$) & $1.58$ & $1.10$ & $0.28$ & $0.28$  &  &   \\
&&&&&&\\
Two-Species  & $1.05$ & $1.6$ & $0.50$ & $0.51$ & $1.25$ & $0.91$ \\
\, Eden Model ($d=1$)  &&&&&\\
\end{tabular}
\end{ruledtabular}
\caption{Comparison of critical exponents for coupled DP and KPZ equations obtained from RG analysis in $d=4-\epsilon$ with uncoupled DP in $d=4-\epsilon$ and $d=1$~\cite{Henkel}. We also include the numerically determined exponents for the two-species Eden model in $d=1$ from Ref.~\cite{Kuhr2011}; noting that the comparison to the trends from
$\epsilon$-expansion is not particularly favorable.}\label{table:exp}
\end{table}

\subsubsection{$\theta$ nonzero and finite sector}

The full set of RG equations are sufficiently complex to make obtaining fixed points with $\theta\neq 0$ quite challenging.
To the best of our knowledge, there are no more stable fixed points.

\subsubsection{Discussion}

Within the perturbative approach here, we have found a single stable fixed point and extracted its scaling behavoir.
To access the phenomenology of the dynamics near this fixed point, we can write the associated Langevin equations by keeping only terms that have nonzero fixed point values:
 \begin{align}\label{eq:fixedPTLang1}
0&= \nabla^2 h +\frac{\alpha}{\nu} f \\  \label{eq:fixedPTLang2}
\partial_t f &= K\nabla^2 f+\tau f-\frac{g}{2}f^2 +\beta\nabla (f\nabla h)+\sqrt{2D_f f}\eta, 
\end{align}
with the important proviso that $\alpha$ and $\beta\equiv\beta_1=\beta_2$ have  opposite signs ($\delta_j <0$).

Interestingly, at this fixed point $\theta=0$, suggesting that the relaxation dynamics of the KPZ equations are very fast $\nu\gg K$ (with $\alpha/\nu$ flowing to a finite value), which we interpret as the height dynamics `instantaneously' relaxing to match the DP order parameter configuration.
Another point of note is that the coupling of the height back into the order parameter enters as an advection term with effective flow field $\nabla h$: this advective term pushes the active wild-type sites to the maxima or minima of surface undulations depending on the sign of $\beta$.
In either case, however, the effects of $\alpha$ and $\beta$ counteract each other.
For example, suppose $\alpha >0$ and $\beta<0$.
The positive $\alpha>0$ would cause regions of dense active wild-type sites to bulge out, creating regions of negative curvature.
However, $\beta<0$ then drives domain boundaries up the slopes condensing the region of active wild-type sites, thereby slowing the surface growth.
In effect, washing out surface undulations.
This picture is consistent with a negative roughening exponent $\chi=-\epsilon/6$ and a flat front.

Given a configuration $f({\bf x},t)$, the profile of the front  
is formally obtained from Eq.~\eqref{eq:fixedPTLang1}  as $h({\bf x},t)=(\nu/\alpha)\nabla^{-2}f$.
Substituting this in Eq.~\eqref{eq:fixedPTLang2} indicates that coupling to  the height  profile results
in a non-local advection of the composition front through a term $(|\alpha\beta|/\nu)\nabla\left(f\nabla^{-1}f\right)$.
The exponents in Table~\ref{table:exp} thus correspond to those of such a field theory.

\section{Summary}

Using symmetry arguments, we have proposed a phenomenological model for the co-evolution of surface roughness and  the 
evolutionary/ecological dynamics of bacterial range expansions through a coupling of the Directed Percolation and KPZ equations.
In addition to the usual parameters of DP and KPZ, our model includes only four new parameters ($\alpha$, $\beta_1$, $\beta_2$,
and $c$) which have well defined macroscopic manifestations. In particular, we have explored how these parameters modify
the profile of a growing film and can in turn be computed through macroscopic measurements.
We have further explored how the coupling of the DP composition to a rough front profile modifies the critical
behavior at the DP transition (in turn modifying the scaling of KPZ fluctuations) through an RG analysis.
In fact all couplings become relevant below the upper critical dimension of four, suggesting a perturbative analysis in
$d=4-\epsilon$, which we carry out. 
Unfortunately (but perhaps not surprisingly) the RG analysis is not conclusive: the most promising candidate exhibits
`weak dynamic scaling,' with exponents that cannot be profitably compared to existing numerical estimates in $d=1$.
We may hope that extensive numerical simulations (including in higher dimensions), along with further analysis of the coupled
equations, can shed more light on this problem.
 
Through various mappings the KPZ equation can be related to the Burgers equation for a randomly stirred fluid and
to meanderings of a directed path in a random medium. Through similar steps, we show that certain limits of our proposed equations
can be mapped (by a generalized Cole-Hopf transformation) to genealogical paths following ancestry of growing/mutating species, 
and that this mapping also enables exact solution of the co-evolving composition and front profile in the noiseless limit.
We expect that variants of these equations are relevant to a host of other problems, including range expansions of chiral~\cite{George2017} or mutualistic~\cite{Lavrentovich2014} bacteria, or even growth in a weakly compressible~\cite{Plummer2018} or turbulant fluid flow~\cite{Antonov2009}.

\acknowledgements
The authors grateful acknowledge conversations with Alexandre Solon, Sherry Chu, and David R. Nelson.
JMH is supported by the Gordon and Betty Moore Foundation as a Physics of Living Systems Fellow through Grant No. GBMF4513.  MK acknowledges support from the NSF through grant number DMR-1708280.

\appendix
 \section{Microscopic dynamics growth rules}\label{sec:Algorithm}

\subsection{Flat growth}

Illustrations with flat growth are implemented using the Domany-Kinzel (DK) cellular automaton rules or equivalently the ``stepping stone'' model~\cite{Korolev2010,Lavrentovich2013}.
Simulations are performed on a lattice $L$ sites wide and $T$ sites deep.
Each lattice site on layer $n\in [1,T]$ and position $i\in[1,L]$ is labeled by a composition value $f_{n,i}$ taking values $-1$ or $1$ for the two bacterial species, pictured respectively as blue or gray, and $0$ on empty sites.

Each new layer is determined sequentially based on the composition of the previous layer.
In updating the $n$-th layer, the identity in each position $i$ is determined by two `neighbors' in the layer below, $n-1$.
To implement the interactions specified by the tilted square lattice depicted in Fig.~\ref{fig:illustration}, the layer-below neighbors for even layers are $i$ and $i+1$, whereas the layer-below neighbors for odd layers are $i-1$ and $i$. 
The composition value at each lattice site $f_{n,i}$ along layer $n$ is then determined probabilistically 
from layer $n-1$ according to the rules
\begin{align}
&P(f_{n,i}=1| f_{n-1,i}= f_{n-1,i+1}=1)=1-\mu_\mu\\
&P(f_{n,i}=-1| f_{n-1,i}= f_{n-1,i+1}=-1)=1\\
&P(f_{n,i}=1| f_{n-1,i}\neq f_{n-1,i+1})=(1+ \tau_\mu)/2,
\end{align}
for $n$ even, with symmetric rules if $n$ is odd.
 Here, $\mu_\mu$ is a microscopic mutation probability from species $1\to -1$, and $\tau_\mu$ is a microscopic selection bias (or growth rate differential) favoring the reproduction of species 1.

\subsection{Rough growth}

Rough growth simulations are implemented with a rough variant of the DK rules, sometimes refereed to as the modified ``brick wall'' solid-on-solid model~\cite{Drossel2000,Drossel2003}, or ``single step'' model~\cite{Meakin1986, Plischke1987}.
As in the the stepping stone model, simulations are performed on a lattice $L$ sites wide and $T$ sites deep.
Each lattice site on layer $n\in [1,T]$ and position $i\in[1,L]$ is labeled by a composition value $f_{n,i}$ taking values $-1$ or $1$ for the two bacterial species, pictured respectively as blue or gray, and $0$ on empty sites.
However, here the composition value of lattice sites are not updated layer by layer, but sites are updated randomly.

Updates proceed by first randomly choosing a position $i$ and then determining the last nonzero layer where the composition value has not been updated, say $n$.
The addition is then accepted depending on whether the two `neighbors' in the layer below are nonzero.
Again, to implement the interactions specified by the tilted square lattice depicted in Fig.~\ref{fig:illustration}, the layer-below neighbors for even layers are $i$ and $i+1$, whereas the layer-below neighbors for odd layers are $i-1$ and $i$. 
If either of the neighbors in the layer below is zero then the addition is rejected.
If they are both full, rejections are composition dependent with probabilities, say for an even layer,  
\begin{align}
&P({\rm rejection}| f_{n-1,i}= f_{n-1,i+1}=1)=0\\
&P({\rm rejection}| f_{n-1,i}\neq f_{n-1,i+1})=\alpha_\mu\\
&P({\rm rejection}| f_{n-1,i}= f_{n-1,i+1}=-1)=2\alpha_\mu.
\end{align}

If the addition of a new site is accepted, the composition is determined in a manner akin to the flat growth DK rules, except allowing for the influence of same level neighbors weighted by a factor $\beta_\mu$.
The algorithm proceeds by first calculating the weighted fraction of neighbors with composition value $1$ of a newly added site $(n,i)$:
\begin{align}
\omega_{n,i}=\frac{\Theta(f_{n-1,i})+\Theta(f_{n-1,i+1})+\beta_\mu(\Theta(f_{n,i-1})+\Theta(f_{n,i+1}))}{|f_{n-1,i}|+|f_{n-1,i+1}|+\beta_\mu(|f_{n,i-1}|+|f_{n,i+1}|)},
\end{align}
where $\Theta(x)$ is the heaviside step function taking value $1$ for $x>0$ and $0$ otherwise.
Then based on this value the composition at site $(n,i)$ is determined from the probability distribution
\begin{align}
&P(f_{n,i}=1|\omega_{n,i}= 1)=1- \mu_\mu\\
&P(f_{n,i}=1|0<\omega_{n,i}< 1)=(1+ \tau_\mu) (1-\mu_\mu)\omega_{n,i}\\
&P(f_{n,i}=1|\omega_{n,i}= 0)=0,
\end{align}
which are well defined probabilities for small $\beta_\mu$,  $\tau_\mu$ and $\mu_\mu$, which is our region of interest.
 Apart from the illustrations in Fig.~\ref{fig:illustration}, all numerical calculations are performed with $\tau_\mu=\mu_\mu=0$.

Updates are performed a fixed number of times. All simulations finish before the addition of any new sites in the last layer $T$.

\begin{widetext}
\section{Renormalization group calculation}\label{sec:RGeqs}

The RG program operates by tracking the effect on the parameters of repeatedly integrating out short wavelengths (high momenta).
However, this is only analytically tractable for nonlinear theories perturbatively around a solvable Gaussian theory near a critical dimension.
Here, the solvable theory is provided by the Gaussian part of the MSRJD action [Eq.~\eqref{eq:Gaus}] with nonlinear terms in the action  [Eq.~\eqref{eq:nonLinear}] treated as small corrections.

To facilitate our presentation, let us introduce the vector of fields $S=({\tilde h}, { h}, {\tilde f}, { f})$.
Furthermore, as we have already argued that $D_h$ is irrelevant under RG while $c$ does not renormalize at
the lowest order, allowing us to ignore them.
That leaves for the Gaussian action [Eq.~\eqref{eq:Gaus}] in Fourier space [with for example $k=(\omega, q)$ and $h(x,t)=\int \frac{d^{d}x d\omega}{(2\pi)^d}\ e^{iqx-i\omega t}h(q,\omega)$]
\begin{equation}
{\mathcal J}_0=\frac{1}{2}\int \frac{d^{d'}k}{(2\pi)^{d'}}\ S(k){\mathcal A}^{-1}(k)S(-k),
\end{equation}
where we have introduced the shorthand $d'=d+1$ and the correlation matrix 
\begin{equation}
\begin{split}
&{\mathcal A}(k)=\\
& \left(\begin{array}{cccc}0 & \frac{1}{i\omega+\nu q^2} & 0 &  \\ \frac{1}{-i\omega+\nu q^2} & 0 & \frac{\alpha}{(-i\omega+\nu q^2)(-i\omega + K q^2-\tau)} & 0 \\0 & \frac{\alpha}{(i\omega+\nu q^2)(i\omega + K q^2-\tau)}  & 0 & \frac{1}{i\omega + K q^2-\tau} \\0 & 0 & \frac{1}{-i\omega + K q^2-\tau} & 0\end{array}\right),
\end{split}
\end{equation}
whose elements determine the unperturbed (connected) correlation functions or cumulants, sometimes called the bare propagators, $\langle S_m(k)S_n(l)\rangle^c=(2\pi)^d{\mathcal A}_{mn}(k)\delta(k+l)$:
\begin{align}
H_0(k)&\equiv\frac{1}{(2\pi)^{d'}}\langle {\tilde h}(k)h(-k)\rangle_0 = \frac{1}{i\omega+\nu q^2}\\
F_0(k)&\equiv\frac{1}{(2\pi)^{d'}}\langle {\tilde f}(k)f(-k)\rangle_0 = \frac{1}{i\omega+K q^2-\tau}\\
G_0(k)&\equiv\frac{1}{(2\pi)^{d'}}\langle {\tilde f}(k)h(-k)\rangle_0 = \frac{\alpha}{(i\omega+\nu q^2)(i\omega+K q^2-\tau)}.\label{eq:fhcontract}
\end{align}
Renormalization proceeds by considering how the behavior of these correlation functions change when one perturbatively includes the short wavelength effects of the nonlinear interactions in the action [Eq.~\eqref{eq:nonLinear}], which in Fourier space read
\begin{equation}
\begin{split}
{\mathcal J}_{\rm int}=\int \frac{d^{d'}k_1d^{d'}k_2d^{d'}k_3}{(2\pi)^{2d'}}\bigg(&\frac{g}{2}{\tilde f}(k_1)f(k_2)f(k_3)-D_f{\tilde f}(k_1){\tilde f}(k_2)f(k_3)+\frac{\lambda}{2}(q_2\cdot q_3){\tilde h}(k_1)h(k_2)h(k_3) \\
&+\beta_1(q_2\cdot q_3) {\tilde f}(k_1)f(k_2)h(k_3)+\beta_2q_3^2 {\tilde f}(k_1)f(k_2)h(k_3)
\bigg)\delta(k_1+k_2+k_3).
\end{split}
\end{equation}

A formally instructive way to organize this perturbation expansion is to consider the generating function for the cumulants of the distribution~\cite{Tauber}
\begin{equation}
{\mathcal Z}[{\tilde j},j]=\ln\left\langle \exp\int d^{d}x\int dt\ \sum_\alpha \left({\tilde j}^\alpha {\tilde S}^\alpha+j^\alpha S^\alpha\right)\right\rangle,
\end{equation}
whose derivates with respect to the dummy variables $j$ and ${\tilde j}$ generate the cumulants.
For example, the two-point correlation function is
\begin{equation}
\begin{split}
{\mathcal G}_{\alpha\beta}(x,t;y,s)&=\left\langle S^\alpha(x,t){\tilde S}^\beta(y,s)\right\rangle_c \\
&=\frac{\delta}{\delta j^\alpha(x,t)}\frac{\delta}{\delta {\tilde j}^\beta(y,s)} {\mathcal Z}\Big|_{{\tilde j}=j=0}.
\end{split}
\end{equation}
The perturbative analysis of the cumulants can be simplified by constructing an alternative set of functions known as 
{\it  vertex functions}~\cite{Tauber}.
They are obtained, like the cumulants, from a generating function as follows.
We first define the conjugate variables to $j$ and ${\tilde j}$
\begin{equation}
{\tilde \Phi}^\alpha(x,t)=\frac{\delta{\mathcal Z}[{\tilde j},j]}{\delta {\tilde j}^\alpha(x,t)},\quad \Phi^\alpha(x,t)=\frac{\delta{\mathcal Z}[{\tilde j},j]}{\delta j^\alpha(x,t)},
\end{equation}
and from them  obtain the vertex generating function as the Legendre transform of the cumulant generating function
\begin{equation}
\Gamma[{\tilde \Phi},\Phi]=-{\mathcal Z}[{\tilde j},j]+\int d^{d}x\int dt\, \sum_\alpha \left({\tilde j}^\alpha {\tilde \Phi}^\alpha+j^\alpha \Phi^\alpha\right).
\end{equation}
Its derivatives at the origin define the vertex functions
\begin{equation}
\Gamma^{(\tilde{N},N)}_{\{\alpha_i\};\{\beta_k\}}(\{x_i,t_i\};\{y_k,s_k\})=\prod_{i=1}^{\tilde N}\frac{\delta}{\delta \tilde\Phi^{\alpha_i}(x_i,t_i)}\prod_{k=1}^{N}\frac{\delta}{\delta \Phi^{\beta_k}(y_k,s_k)} \Gamma[\tilde\Phi,\Phi]\bigg|_{j={\tilde j}=0}.
\end{equation}
Their relationship to the cumulants can be obtained by successive derivatives.
For example, the first derivatives are
\begin{equation}
\frac{\delta \Gamma[\tilde\Phi,\Phi]}{\delta \tilde\Phi^\alpha(x,t)}={\tilde j}^\alpha(x,t),\qquad\qquad\frac{\delta \Gamma[\tilde\Phi,\Phi]}{\delta \Phi^\beta(x,t)}=j^\beta(x,t) ,
\end{equation}
as is expected for a Legendre transform.
Another important example is obtained from two derivatives
\begin{equation}
\Gamma^{(1,1)}_{\alpha,\beta}(q,\omega)={\mathcal G}^{-1}_{\alpha,\beta}(-q,-\omega),
\end{equation}
demonstrating that the two point vertex is the matrix inverse of the cumulants (or matrix of propagators).
Successive derivatives lead to higher-order vertices.
Importantly, as we turn to the perturbative calculation, the corrections to the vertex functions are determined from the class of one-particle irreducible diagrams.

With this formalism in hand, we now calculate the corrections to the vertex functions due to the effects of the short wavelength behavior propagated through the nonlinearities.
This proceeds by perturbatively expanding the vertex function in the nonlinearities and then integrating over a small annulus of short wavelengths, $\Lambda/b<q<\Lambda$, with $b\gtrsim1$.
We will see that this perturbation theory is consistent -- that is all the nonlinearities are indeed small -- only near the critical dimension $d=4-\epsilon$.
In addition, the critical value of the control parameter near $d=4-\epsilon$ is also expected to be small, allowing us to simplify the presentation by noting that within the order of our perturbation expansion $\tau$ will not affect the renormalization of any other parameter, but itself.
With this in mind, we consider the lowest order perturbative corrections to the vertex functions, which can be graphically represented using Feynman diagrams containing one loop.

The first vertex function we consider is the inverse of the height response function
\begin{equation}
\Gamma_{{\tilde h},h}^{(1,1)}(\omega, q)=\frac{1}{{\mathcal G}_{\tilde{h},h}(-\omega,-q)}\approx-i\omega +\nu q^2,
\end{equation}
which has no corrections to one-loop order.
Thus, $\nu$ does not renormalize.
(Note that in the standard KPZ equation, there is renormalization of $\nu$ due to finite $D_h$, which is absent here.)

Next, we consider the vertex function corresponding to the inverse of the DP response function, whose corrections are listed in Fig.~\ref{fig:fPropogator}:
\begin{align}
\Gamma^{(1,1)}_{{\tilde f}f}=\frac{1}{{\mathcal G}_{\tilde{f}f}(-\omega,-q)}&\approx-i\omega+Kq^2-\tau+(a)+ (b),
\end{align}
where
\begin{figure}
\includegraphics[scale=.6]{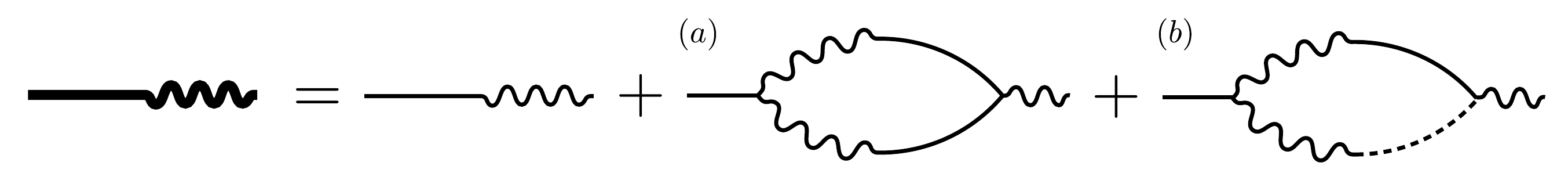}
\caption{Diagrammatic expansion for the propagator $\Gamma^{(1,1)}_{{\tilde f},f}$:
the field $f$ is represented by a straight line and $\tilde{f}$ by a wavy line.
The correction (\emph{a}) is proportional to the product $D_f$ (left vertex) and $g$ (right vertex), 
while (\emph{b}) arises from the product of $D_f$ and $\beta_1,\beta_2$ (right vertex).
Note that the contraction of $h$ (dashed line) and $\tilde{f}$ brings a factor of $\alpha$
as indicated in Eq.~\eqref{eq:fhcontract}.}\label{fig:fPropogator}
\end{figure}
\begin{align}
(a) &=gD_f\int \frac{d^{d'}k^\prime}{(2\pi)^{d'}}F_0(k-k^\prime)F_0(k^\prime)=\frac{gD_fA_{4}}{2K}\left(\Lambda^2-\frac{-i\omega+Kq^2/2}{2K}\right)\left(1-\frac{1}{b}\right),\\
(b) &= 2D_f\int \frac{d^{d'}k^\prime}{(2\pi)^{d'}}F_0(k-k^\prime)G_0(k^\prime)\left[\beta_1q'\cdot(q-q')+\beta_2q'^2\right] \\
\nonumber
&= \alpha D_f A_{4}\bigg\{\frac{(\beta_2-\beta_1)\Lambda^2}{K(K+\nu)}+i\omega\frac{(\beta_2-\beta_1)(3K+\nu)}{2K^2(K+\nu)^2}\\
&\qquad +\frac{q^2}{K(K+\nu)}\left[\beta_1\frac{K^2+4K\nu+\nu^2}{2(K+\nu)^2}+\beta_2\frac{K^2-4K\nu-\nu^2}{4(K+\nu)^2}\right]
\bigg\}\left(1-\frac{1}{b}\right),
\end{align}
where integrals were performed by first expanding to first order in $\omega$ and $q$, since we are only interested in the long time, long wavelength behavior.
The  procedure for performing such integrals has been described previously, for example in the Appendix of Ref.~\cite{Medina1989}.

The next two-point vertex to renormalize comes from the coupling of the response field ${\tilde h}$ to the DP composition field $f$.
Its perturbative expansion depicted in Fig.~\ref{fig:alphaProp}, corresponds to
\begin{equation}
\Gamma^{(1,1)}_{{\tilde h}f}(q,\omega)=-\frac{{\mathcal G}_{{\tilde f}h}(-q,-\omega)}{{\mathcal G}_{{\tilde f} f}(-q,-\omega){\mathcal G}_{{\tilde h}h}(-q,-\omega)}\approx-\alpha+(a),
\end{equation}
\begin{figure}
\includegraphics[scale=.6]{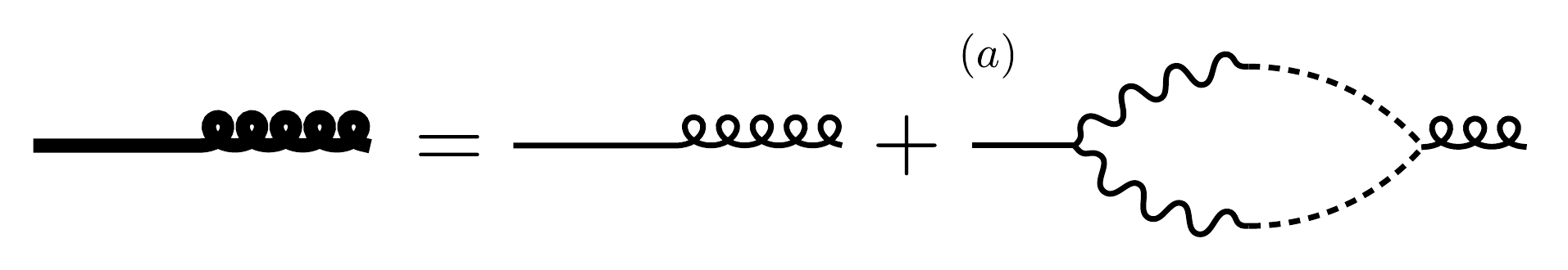}
\caption{Diagrammatic expansion for the propagator  $\Gamma^{(1,1)}_{{\tilde h},f}$, with $\tilde{h}$ 
represented by the curly line. The correction in (\emph{a}) is proportional to the
product of $D_f$ (left vertex) and $\lambda$ (right vertex).}\label{fig:alphaProp}
\end{figure}
where the only correction surviving with the external momenta set to zero $q=\omega=0$ is
\begin{align}
(a) = -\lambda D_f\int \frac{d^{d'}k^\prime}{(2\pi)^{d'}}\ q'^2G_0(-k^\prime)G_0(k^\prime)=-\frac{\lambda D_f\alpha^2A_4}{2\nu K(K+\nu)}\left(1-\frac{1}{b}\right).
\end{align}

There are four three-point vertices corresponding to each of the nonlinear parameters:
\begin{itemize}

\item
We first address the KPZ nonlinearity $\lambda$ whose renormalization corresponds to the vertex function
\begin{equation}
\Gamma_{{\tilde h},{\tilde h},h}^{(2,1)}\approx-\lambda,
\end{equation}
with no corrections at one-loop order, implying that $\lambda$ does not renormalize at this order.
\item
On the other hand, the DP nonlinearities both renormalize.
The corrections to the noise vertex
\begin{equation}
\Gamma_{{\tilde f}{\tilde f}f}^{(2,1)}\approx 2D_f+(a)+(b)+(c),
\end{equation}
depicted in Fig.~\ref{fig:D}
\begin{figure}
\includegraphics[scale=.6]{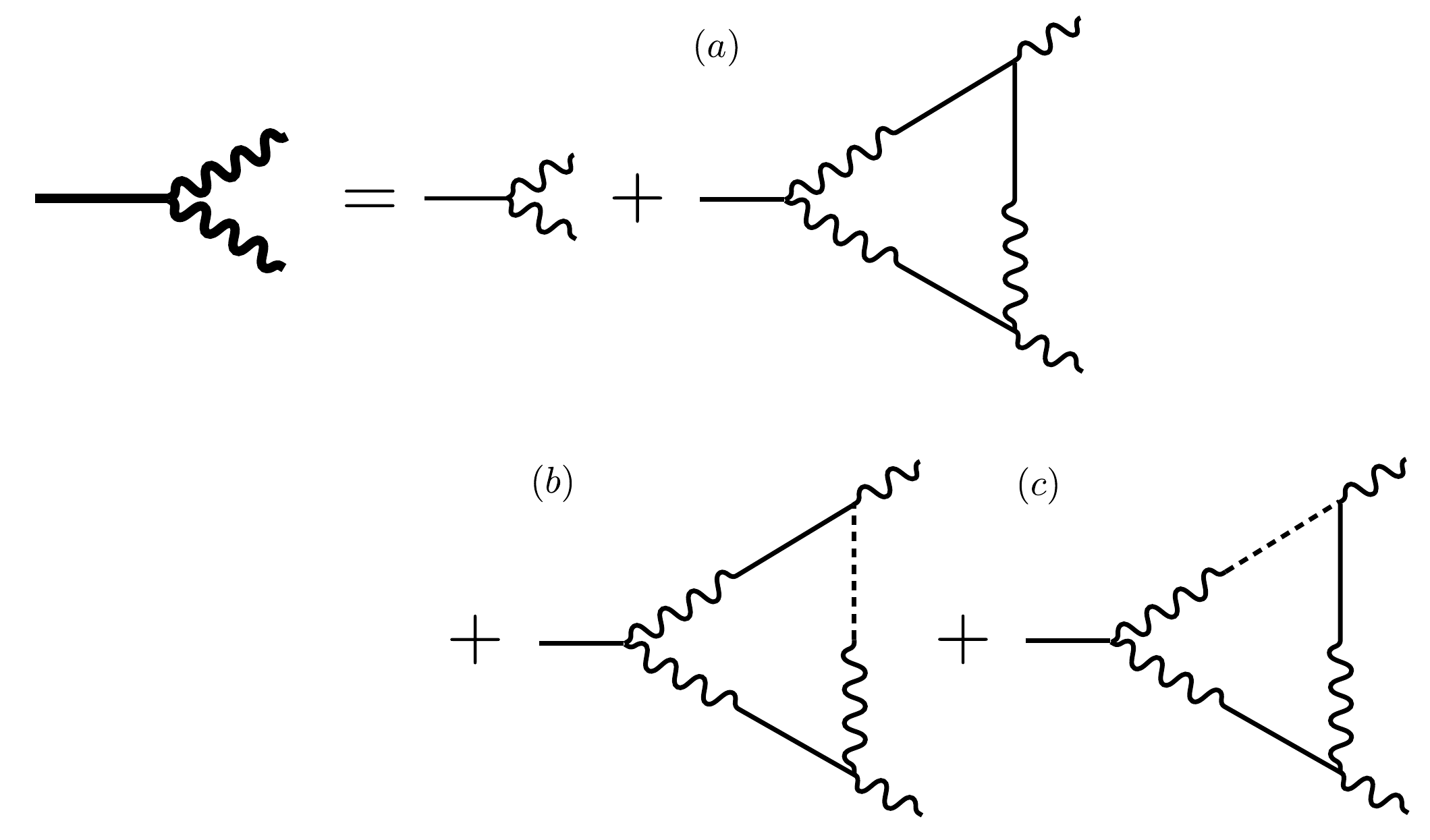}
\caption{Diagrammatic expansion for the vertex function $\Gamma^{(2,1)}_{{\tilde f},{\tilde f},f}$.}\label{fig:D}
\end{figure}
correspond to the integrals evaluated at zero external momenta
\begin{align}
(a) &= -8 g D^2_f\int \frac{d^{d'}k^\prime}{(2\pi)^{d'}}\ F_0(k^\prime)F_0(-k^\prime)F_0(-k^\prime)=-\frac{2gD_f^2A_4}{K^2}\left(1-\frac{1}{b}\right)\\
(b) &= -8 D^2_f\int \frac{d^{d'}k^\prime}{(2\pi)^{d'}}\ F_0(k^\prime)F_0(-k^\prime)G_0(-k^\prime)\left[\beta_2q'^2-\beta_1 q
'^2\right] =-\frac{2\alpha(\beta_2-\beta_1)D_f^2A_4}{K^2(K+\nu)}\left(1-\frac{1}{b}\right)\\
(c) &= -8 D^2_f\int \frac{d^{d'}k^\prime}{(2\pi)^{d'}}\ G_0(k^\prime)F_0(-k^\prime)F_0(-k^\prime)\left[\beta_2q'^2-\beta_1 q
'^2\right] =-2\alpha(\beta_2-\beta_1)D_f^2A_4\frac{3K+\nu}{K^2(K+\nu)^2}\left(1-\frac{1}{b}\right).
\end{align}
\item
The other DP nonlinearity's renormalization is captured with the vertex 
\begin{equation}
\Gamma_{{\tilde f}ff}^{(1,2)}\approx-g+(a)+(b)+(c)+(d)+(e)+(f),
\end{equation}
where the Feynman diagrams for the corrections are in Fig.~\ref{fig:g}. 
\begin{figure}
\includegraphics[scale=.5]{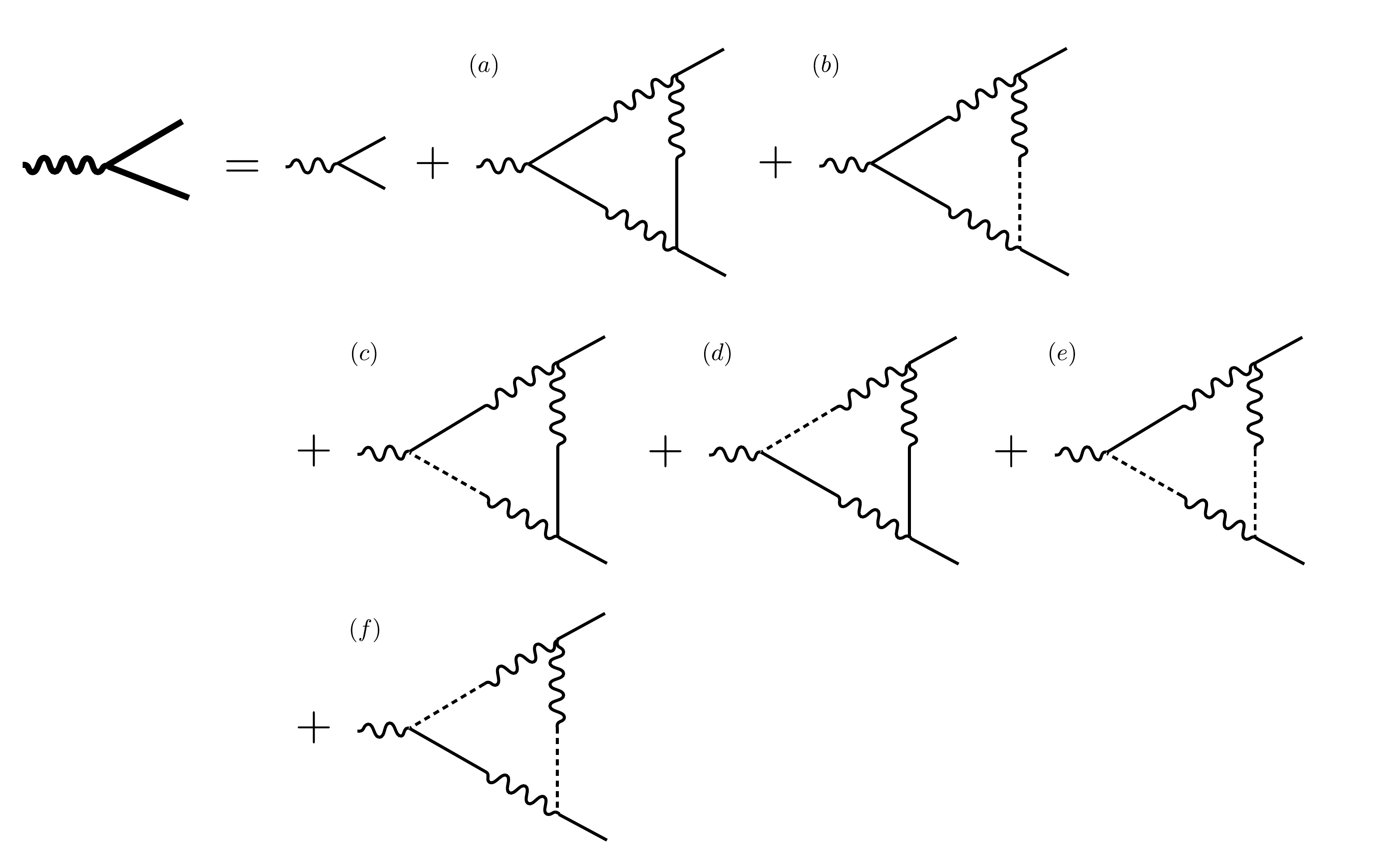}
\caption{Diagrammatic expansion for the vertex function $\Gamma^{(1,2)}_{{\tilde f},f,f}$.}\label{fig:g}
\end{figure}
The corresponding integrals at zero external momenta are
\begin{align}
(a) &= 4 g^2 D_f\int \frac{d^{d'}k^\prime}{(2\pi)^{d'}}\ F_0(k^\prime)F_0(-k^\prime)F_0(-k^\prime)=\frac{g^2D_fA_4}{K^2}\left(1-\frac{1}{b}\right)\\
(b) &= 4 g D_f\int \frac{d^{d'}k^\prime}{(2\pi)^{d'}}\ \beta_2q'^2 F_0(k^\prime)F_0(-k^\prime)G_0(-k^\prime)=\frac{gD_f\alpha \beta_2 A_4}{K^2(K+\nu)}\left(1-\frac{1}{b}\right)\\
(c) &= 4 g D_f\int \frac{d^{d'}k^\prime}{(2\pi)^{d'}}\ F_0(k^\prime)F_0(-k^\prime)G_0(-k^\prime)\left[\beta_2q'^2-\beta_1 q
'^2\right] =\frac{gD_f\alpha (\beta_2-\beta_1) A_4}{K^2(K+\nu)}\left(1-\frac{1}{b}\right)\\
(d) &= 4 g D_f\int \frac{d^{d'}k^\prime}{(2\pi)^{d'}}\ F_0(k^\prime)F_0(-k^\prime)G_0(-k^\prime)\left[\beta_2q'^2-\beta_1 q
'^2\right] =gD_f\alpha (\beta_2-\beta_1) A_4\frac{3K+\nu}{K^2(K+\nu)^2}\left(1-\frac{1}{b}\right)\\
(e) & = 2g D_f\int \frac{d^{d'}k^\prime}{(2\pi)^{d'}}\ G_0(k^\prime)F_0(-k^\prime)G_0(-k^\prime)\left[\beta_2q'^2-\beta_1 q
'^2\right] =\frac{\alpha^2D_fA_4\beta_2(\beta_2-\beta_1)}{K^2(K+\nu)^2}\left(1-\frac{1}{b}\right)\\
(f) &= 4 D_f\int \frac{d^{d'}k^\prime}{(2\pi)^{d'}}\ G_0(k^\prime)G_0(-k^\prime)F_0(-k^\prime)\left[\beta_2q'^2-\beta_1 q
'^2\right] \beta_2q'^2 =\alpha^2 D_f\beta_2(\beta_2-\beta_1)A_4\frac{2(K+\nu)}{K^2\nu(K+\nu)^2}\left(1-\frac{1}{b}\right)
\end{align}
\item
Finally, the nonlinear coupling of the height field to the DP order parameter occurs through the two beta couplings, whose renormalization can be handled together through the vertex
\begin{equation}
\Gamma_{{\tilde f}fh}^{(1,2)}\approx -\beta_1(q_1\cdot q_2)-\beta_2 q_1^2 +(a)+(b)+(c)+(d),
\end{equation}
where the corrections, depicted in Fig.~\ref{fig:betaRG}, are
\begin{figure}
\includegraphics[scale=.5]{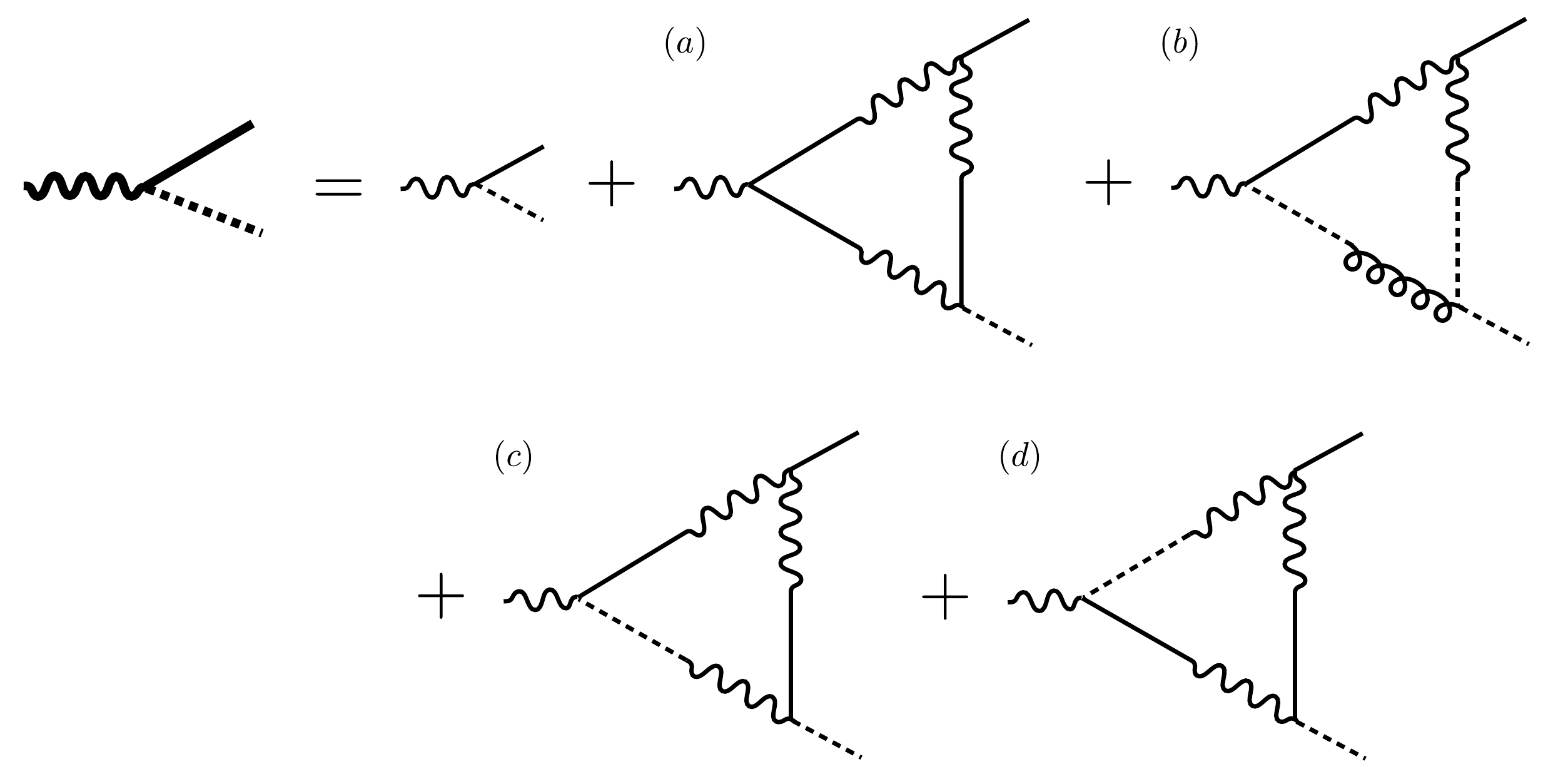}
\caption{Diagrammatic expansion for the vertex function $\Gamma_{{\tilde f}fh}^{(1,2)}$.}\label{fig:betaRG}
\end{figure}
\begin{align}
\nonumber
(a) &= 2g D_f\int \frac{d^{d'}k^\prime}{(2\pi)^{d'}}\ F_0\left(\frac{k_1+k_2}{2}+k^\prime\right)F_0\left(\frac{k_1+k_2}{2}-k^\prime\right)F_0\left(\frac{k_1-k_2}{2}-k^\prime\right)\left[\beta_1 q_2\cdot\left(\frac{q_1-q_2}{2}-q'\right)+\beta_2q_2'^2\right] \\
&=\frac{gD_fA_4}{2K^2}\left[q_2^2\left(\beta_2-\frac{\beta_1}{2}\right)+q_1\cdot q_2\frac{\beta_1}{2}\right]\left(1-\frac{1}{b}\right)\\
\nonumber
(b) &= 2\lambda D_f\int \frac{d^{d'}k^\prime}{(2\pi)^{d'}}\ F_0\left(\frac{k_1+k_2}{2}+k^\prime\right)H_0\left(\frac{k_1+k_2}{2}-k^\prime\right)G_0\left(\frac{k_1-k_2}{2}-k^\prime\right) \\
\nonumber
&\hspace{.1\textwidth} \times\left[q_2\cdot \left(\frac{q_1-q_2}{2}-q'\right)\right]\left[\beta_1 \left(\frac{q_1+q_2}{2}+q'\right)\cdot\left(\frac{q_1+q_2}{2}-q'\right)+\beta_2\left(\frac{q_1+q_2}{2}-q'\right)^2\right] \\
&=\frac{\alpha\lambda D_fA_4}{K(K+\nu)^3}\left[q_2^2\left(\beta_2\frac{K}{2}-\beta_1\frac{K-\nu}{4}\right)+q_1\cdot q_2\left(\beta_2\frac{5K+\nu}{4}-\beta_1\frac{K-\nu}{4}\right)\right]\left(1-\frac{1}{b}\right)\\
\nonumber
(c) &= 2 D_f\int \frac{d^{d'}k^\prime}{(2\pi)^{d'}}\ F_0\left(\frac{k_1+k_2}{2}+k^\prime\right)G_0\left(\frac{k_1+k_2}{2}-k^\prime\right)F_0\left(\frac{k_1-k_2}{2}-k^\prime\right) \\
\nonumber
&\hspace{.1\textwidth} \times\left[\beta_1 \left(\frac{q_1+q_2}{2}+q'\right)\cdot\left(\frac{q_1+q_2}{2}-q'\right)+\beta_2\left(\frac{q_1+q_2}{2}-q'\right)^2\right] \left[\beta_1 \left(\frac{q_1-q_2}{2}-q'\right)\cdot q_2+\beta_2q_2^2\right] \\
&=\frac{\alpha D_fA_4}{2K^2(K+\nu)^2}\left[q_2^2\left(\beta^2_2-\beta_1\beta_2\frac{4K+5\nu}{4(K+\nu)}+\beta^2_1\frac{\nu}{2(K+\nu)}\right)+q_1\cdot q_2\left(\beta_1\beta_2 \frac{2K+\nu}{2(K+\nu)}-\beta^2_1\frac{3K+\nu}{4(K+\nu)}\right)\right]\left(1-\frac{1}{b}\right)\\
\nonumber
(d) &= 2 D_f\int \frac{d^{d'}k^\prime}{(2\pi)^{d'}}\ G_0\left(\frac{k_1+k_2}{2}+k^\prime\right)F_0\left(\frac{k_1+k_2}{2}-k^\prime\right)F_0\left(\frac{k_1-k_2}{2}-k^\prime\right) \\
\nonumber
&\hspace{.1\textwidth} \times\left[\beta_1 \left(\frac{q_1+q_2}{2}+q'\right)\cdot\left(\frac{q_1+q_2}{2}-q'\right)+\beta_2\left(\frac{q_1+q_2}{2}+q'\right)^2\right] \left[\beta_1 \left(\frac{q_1-q_2}{2}-q'\right)\cdot q_2+\beta_2q_2^2\right] \\
&=\frac{\alpha D_fA_4}{2K^2(K+\nu)^2}\bigg[q_2^2\left(\beta^2_2(3K+\nu)-\beta_1\beta_2\frac{41K^2+52K\nu+13\nu^2}{2(K+\nu)}+\beta^2_1\frac{K^2+K\nu+\nu^2}{K+\nu}\right) \\\nonumber
&\qquad\qquad\qquad\qquad\qquad-q_1\cdot q_2\left(\beta_1\beta_2 \frac{K^2-4K\nu-\nu^2}{2(K+\nu)}+\beta^2_1\frac{K^2+12K\nu+3\nu^2}{4(K+\nu)}\right)\bigg]\left(1-\frac{1}{b}\right)
\end{align}
\end{itemize}

Next, we rescale space, time and the fields as $x\to b x$, $t\to b^z t$, $f\to b^{y_f}f$, $h\to b^{y_h}h$, ${\tilde f}\to b^{{\tilde y}_f}{\tilde f}$, and ${\tilde h}\to b^{{\tilde y}_h}{\tilde h}$.
The corrections are then absorbed into the model parameters, so that the formal structure of the theory remains invariant.
The new effective model parameters now have changed by a magnitude that depends on $b$.
By letting $b=e^l$, with $l\ll1$, and setting $\Lambda=1$ for convenience, the change (or flow) in these parameters is described by the collection of RG equations
\begin{align}
&\frac{d\nu}{dl}=(d^\prime +z-2+{\tilde y}_h+y_h)\nu=(z-2)\nu\\
&\frac{d\lambda}{dl}=(d^\prime+z-2+{\tilde y}_h+2y_h)\lambda=(z-2+y_h)\lambda\\
&\frac{d\tau}{dl}=(d^\prime+z+{\tilde y}_f+y_f)\tau-gD_f\frac{A_{4}}{2K}-\alpha D_f(\beta_2-\beta_1) \frac{ A_{4}}{K(K+\nu)}\\
&\frac{dK}{dl}=\left(d^\prime +z-2+{\tilde y}_f +y_f-gD_f\frac{A_{4}}{8K^2}\right)K+\alpha D_f\frac{ A_{4}}{K(K+\nu)}\Bigg[\beta_1\frac{K^2+4K\nu+\nu^2}{2(K+\nu)^2}+\beta_2\frac{K^2-4K\nu-\nu^2}{4(K+\nu)^2}\Bigg]\\
&\frac{d\alpha}{dl}=(d^\prime +z+ {\tilde y}_h+y_f)\alpha+\frac{\lambda D_f\alpha^2}{2\nu}\frac{A_{4}}{K(K+\nu)}\\
&\frac{dD_f}{dl}=(d^\prime+z+2{\tilde y}_f+y_f)D_f-\frac{gD_f^2}{K^2}A_4-2\alpha (\beta_2-\beta_1) D_f^2A_4\frac{2K+\nu}{K^2(K+\nu)^2}\\
&\frac{dg}{dl}=(d^\prime+z+{\tilde y}_f+2y_f)g-\frac{g^2D_f}{K^2}A_4+2\alpha g D_f\beta_1A_4\frac{2K+\nu}{K^2(K+\nu)^2}-\alpha g D_f\beta_2A_4\frac{5K+3\nu}{K^2(K+\nu)^2}\nonumber\\ 
&\qquad -2\alpha^2D_f\beta_2(\beta_2-\beta_1) A_4\frac{1}{K^2\nu(K+\nu)}\\
&\frac{d\beta_1}{dl}=(d^\prime+z-2+{\tilde y}_f+y_f+y_h)\beta_1-\frac{gD_f\beta_1}{4K^2}A_4+\alpha\lambda D_f\beta_1\frac{A_4}{(K+\nu)^3}-\alpha\lambda D_f\beta_2A_4\frac{5K+\nu}{4K(K+\nu)^3} \nonumber\\ 
&\qquad+\alpha D_f \beta_1^2A_4\frac{K^2+4K\nu+\nu^2}{2K^2(K+\nu)^3}-\alpha \beta_1\beta_2 D_fA_4\frac{K^2+7K\nu+2\nu^2}{4K^2(K+\nu)^3}\\
&\frac{d\beta_2}{dl}=(d^\prime+z-2+{\tilde y}_f+y_f+y_h)\beta_2+\frac{gD_f\beta_1}{4K^2}A_4-\frac{gD_f\beta_2}{2K^2}A_4+\alpha\lambda D_f\beta_1A_4\frac{K-\nu}{4K(K+\nu)^3} -\alpha\lambda D_f\beta_2A_4\frac{1}{2(K+\nu)^3}\nonumber\\ 
&\qquad-\alpha D_f \beta_1^2A_4\frac{2K^2+K\nu+\nu^2}{4K^2(K+\nu)^3}+\alpha \beta_1\beta_2 D_fA_4\frac{11K^2+14K\nu+5\nu^2}{4K^2(K+\nu)^3}-\alpha D_f \beta_2^2A_4\frac{2K+\nu}{K^2(K+\nu)^2},
\end{align}
coupled with the algebraic constrains on the scaling exponents
\begin{align}
&d^\prime+{\tilde y}_h+y_h=0\\
&d^\prime+{\tilde y}_f+y_f-gD_f\frac{A_{4}}{4K^2}-\alpha D_f(\beta_2-\beta_1) A_{4}\frac{(3K+\nu)}{2K^2(K+\nu)^2}=0,
\end{align}
which come from demanding the terms proportional to $\omega$ in the propagators are scale invariant.

A close examination of the RG equations reveals that they only depend on particular collections of variables, the dimensionless effective parameters in Eq.~(\ref{eq:effPara}).
The associated RG equations for these effective parameters are
\begin{align}
&\frac{du}{dl}=u\left\{\epsilon +\frac{1}{2(1+\theta)^3}\left[-3u+\delta_1(5+12\theta +11\theta^2)-\delta_2(8+20\theta+16\theta^2)\right]\right\}+\frac{2(\delta_1-\delta_2)\delta_2}{(1+\theta)}\\
&\frac{d\zeta}{dl}=\zeta\left\{\epsilon +\frac{\zeta}{8(1+\theta)^3}\left[-5u(1+\theta)^3+4\zeta(1+\theta)^2+\delta_1(8+16\theta+16\theta^2)-2\delta_2(10+24\theta+22\theta^2)\right]\right\}\\
\nonumber
&\frac{d\delta_1}{dl}=\delta_1\bigg\{\epsilon+\frac{1}{4(1+\theta)^3}\left[-u(3-9\theta+9\theta^2+3\theta^3)+\zeta(2+8\theta+2\theta^2)+\delta_1(4+8\theta+8\theta^2)-\delta_2(6+15\theta+13\theta^2)\right]\bigg\} \\
&\qquad\quad-\frac{\zeta\delta_2(1+5\theta)}{4(1+\theta)^3}\\
\nonumber
&\frac{d\delta_2}{dl}=\delta_2\left\{\epsilon+\frac{1}{4(1+\theta)^3}\left[-4u(1+\theta)^3+2\zeta(1+\theta+\theta^2)+\delta_1(7+14\theta+17\theta^2)-\delta_2(8+20\theta+20\theta^2)\right]\right\} \\
&\qquad\quad-\frac{1}{4(1+\theta)^3}\left\{\delta_1^2(1+\theta-2\theta^2)+\delta_1[\zeta(1-\theta)-u(1+\theta)^3]\right\}\\
&\frac{d\theta}{dl}=\frac{\theta}{8}\left\{u+\frac{2}{(1+\theta)^3}\left[\delta_2(1+4\theta+7\theta^2)-4\delta_1\theta^2\right]\right\} \\
&\frac{d\bar\tau}{dl}=\bar\tau\left\{2+\frac{u}{8}-4\delta_1(1+4\theta+\theta^2)+2\delta_2(1+4\theta-\theta^2)\right\}-\frac{u}{2}+\frac{\delta_1-\delta_2}{2(1+\theta)}.
\end{align}
\end{widetext}

\bibliography{DirectedPercolation.bib}

\begin{thebibliography}{54}%
\makeatletter
\providecommand \@ifxundefined [1]{%
 \@ifx{#1\undefined}
}%
\providecommand \@ifnum [1]{%
 \ifnum #1\expandafter \@firstoftwo
 \else \expandafter \@secondoftwo
 \fi
}%
\providecommand \@ifx [1]{%
 \ifx #1\expandafter \@firstoftwo
 \else \expandafter \@secondoftwo
 \fi
}%
\providecommand \natexlab [1]{#1}%
\providecommand \enquote  [1]{``#1''}%
\providecommand \bibnamefont  [1]{#1}%
\providecommand \bibfnamefont [1]{#1}%
\providecommand \citenamefont [1]{#1}%
\providecommand \href@noop [0]{\@secondoftwo}%
\providecommand \href [0]{\begingroup \@sanitize@url \@href}%
\providecommand \@href[1]{\@@startlink{#1}\@@href}%
\providecommand \@@href[1]{\endgroup#1\@@endlink}%
\providecommand \@sanitize@url [0]{\catcode `\\12\catcode `\$12\catcode
  `\&12\catcode `\#12\catcode `\^12\catcode `\_12\catcode `\%12\relax}%
\providecommand \@@startlink[1]{}%
\providecommand \@@endlink[0]{}%
\providecommand \url  [0]{\begingroup\@sanitize@url \@url }%
\providecommand \@url [1]{\endgroup\@href {#1}{\urlprefix }}%
\providecommand \urlprefix  [0]{URL }%
\providecommand \Eprint [0]{\href }%
\providecommand \doibase [0]{http://dx.doi.org/}%
\providecommand \selectlanguage [0]{\@gobble}%
\providecommand \bibinfo  [0]{\@secondoftwo}%
\providecommand \bibfield  [0]{\@secondoftwo}%
\providecommand \translation [1]{[#1]}%
\providecommand \BibitemOpen [0]{}%
\providecommand \bibitemStop [0]{}%
\providecommand \bibitemNoStop [0]{.\EOS\space}%
\providecommand \EOS [0]{\spacefactor3000\relax}%
\providecommand \BibitemShut  [1]{\csname bibitem#1\endcsname}%
\let\auto@bib@innerbib\@empty
\bibitem [{\citenamefont {Henkel}\ \emph {et~al.}()\citenamefont {Henkel},
  \citenamefont {Hinrichsen},\ and\ \citenamefont {L\"{u}beck}}]{Henkel}%
  \BibitemOpen
  \bibfield  {author} {\bibinfo {author} {\bibfnamefont {M.}~\bibnamefont
  {Henkel}}, \bibinfo {author} {\bibfnamefont {H.}~\bibnamefont {Hinrichsen}},
  \ and\ \bibinfo {author} {\bibfnamefont {S.}~\bibnamefont {L\"{u}beck}},\
  }\href@noop {} {\emph {\bibinfo {title} {Non-Equilibrium Phase Transitions,
  Volume 1: Absorbing Phase Transitions}}},\ Theoretical and Mathematical
  Physics\ (\bibinfo  {publisher} {Spring Science + Business Media B.\ V.\
  UK})\BibitemShut {NoStop}%
\bibitem [{\citenamefont {\'{O}dor}(2004)}]{Odor2004}%
  \BibitemOpen
  \bibfield  {author} {\bibinfo {author} {\bibfnamefont {G.}~\bibnamefont
  {\'{O}dor}},\ }\href@noop {} {\bibfield  {journal} {\bibinfo  {journal} {Rev.
  Mod. Phys.}\ }\textbf {\bibinfo {volume} {76}},\ \bibinfo {pages} {663 }
  (\bibinfo {year} {2004})}\BibitemShut {NoStop}%
\bibitem [{\citenamefont {T\"{a}uber}(2014)}]{Tauber}%
  \BibitemOpen
  \bibfield  {author} {\bibinfo {author} {\bibfnamefont {U.~C.}\ \bibnamefont
  {T\"{a}uber}},\ }\href@noop {} {\emph {\bibinfo {title} {Critical Dynamics: A
  field theory approach to equilibrium and non-equilibrium scaling behavior}}}\
  (\bibinfo  {publisher} {Cambridge University Press, UK},\ \bibinfo {year}
  {2014})\BibitemShut {NoStop}%
\bibitem [{\citenamefont {Janssen}\ and\ \citenamefont
  {T\"{a}uber}(2005)}]{Janssen2005}%
  \BibitemOpen
  \bibfield  {author} {\bibinfo {author} {\bibfnamefont {H.-K.}\ \bibnamefont
  {Janssen}}\ and\ \bibinfo {author} {\bibfnamefont {U.~C.}\ \bibnamefont
  {T\"{a}uber}},\ }\href@noop {} {\bibfield  {journal} {\bibinfo  {journal}
  {Ann. Phys.}\ }\textbf {\bibinfo {volume} {315}},\ \bibinfo {pages} {147}
  (\bibinfo {year} {2005})}\BibitemShut {NoStop}%
\bibitem [{\citenamefont {T\"{a}uber}\ \emph {et~al.}(2005)\citenamefont
  {T\"{a}uber}, \citenamefont {Howard},\ and\ \citenamefont
  {Vollmayr-Lee}}]{Tauber2005}%
  \BibitemOpen
  \bibfield  {author} {\bibinfo {author} {\bibfnamefont {U.~C.}\ \bibnamefont
  {T\"{a}uber}}, \bibinfo {author} {\bibfnamefont {M.}~\bibnamefont {Howard}},
  \ and\ \bibinfo {author} {\bibfnamefont {B.}~\bibnamefont {Vollmayr-Lee}},\
  }\href@noop {} {\bibfield  {journal} {\bibinfo  {journal} {J. Phys. A}\
  }\textbf {\bibinfo {volume} {38}},\ \bibinfo {pages} {R79 } (\bibinfo {year}
  {2005})}\BibitemShut {NoStop}%
\bibitem [{\citenamefont {Hinrichsen}(2000)}]{Hinrichsen2000}%
  \BibitemOpen
  \bibfield  {author} {\bibinfo {author} {\bibfnamefont {H.}~\bibnamefont
  {Hinrichsen}},\ }\href@noop {} {\bibfield  {journal} {\bibinfo  {journal}
  {Adv. Phys.}\ }\textbf {\bibinfo {volume} {49}},\ \bibinfo {pages} {815}
  (\bibinfo {year} {2000})}\BibitemShut {NoStop}%
\bibitem [{\citenamefont {Korolev}\ \emph {et~al.}(2010)\citenamefont
  {Korolev}, \citenamefont {Avlund}, \citenamefont {Hallatschek},\ and\
  \citenamefont {Nelson}}]{Korolev2010}%
  \BibitemOpen
  \bibfield  {author} {\bibinfo {author} {\bibfnamefont {K.~S.}\ \bibnamefont
  {Korolev}}, \bibinfo {author} {\bibfnamefont {M.}~\bibnamefont {Avlund}},
  \bibinfo {author} {\bibfnamefont {O.}~\bibnamefont {Hallatschek}}, \ and\
  \bibinfo {author} {\bibfnamefont {D.~R.}\ \bibnamefont {Nelson}},\
  }\href@noop {} {\bibfield  {journal} {\bibinfo  {journal} {Rev. Mod. Phys.}\
  }\textbf {\bibinfo {volume} {82}},\ \bibinfo {pages} {1691} (\bibinfo {year}
  {2010})}\BibitemShut {NoStop}%
\bibitem [{\citenamefont {Korolev}\ and\ \citenamefont
  {Nelson}(2011)}]{Korolev2011}%
  \BibitemOpen
  \bibfield  {author} {\bibinfo {author} {\bibfnamefont {K.~S.}\ \bibnamefont
  {Korolev}}\ and\ \bibinfo {author} {\bibfnamefont {D.~R.}\ \bibnamefont
  {Nelson}},\ }\href@noop {} {\bibfield  {journal} {\bibinfo  {journal} {Phys.
  Rev. Lett.}\ }\textbf {\bibinfo {volume} {107}},\ \bibinfo {pages} {088103}
  (\bibinfo {year} {2011})}\BibitemShut {NoStop}%
\bibitem [{\citenamefont {Hallatschek}\ \emph {et~al.}(2007)\citenamefont
  {Hallatschek}, \citenamefont {Hersen}, \citenamefont {Ramanathan},\ and\
  \citenamefont {Nelson}}]{Hallatshcek2007}%
  \BibitemOpen
  \bibfield  {author} {\bibinfo {author} {\bibfnamefont {O.}~\bibnamefont
  {Hallatschek}}, \bibinfo {author} {\bibfnamefont {P.}~\bibnamefont {Hersen}},
  \bibinfo {author} {\bibfnamefont {S.}~\bibnamefont {Ramanathan}}, \ and\
  \bibinfo {author} {\bibfnamefont {D.~R.}\ \bibnamefont {Nelson}},\
  }\href@noop {} {\bibfield  {journal} {\bibinfo  {journal} {PNAS}\ }\textbf
  {\bibinfo {volume} {104}},\ \bibinfo {pages} {19926} (\bibinfo {year}
  {2007})}\BibitemShut {NoStop}%
\bibitem [{\citenamefont {Hallatschek}\ and\ \citenamefont
  {Nelson}(2009)}]{Hallatschek2010}%
  \BibitemOpen
  \bibfield  {author} {\bibinfo {author} {\bibfnamefont {O.}~\bibnamefont
  {Hallatschek}}\ and\ \bibinfo {author} {\bibfnamefont {D.~R.}\ \bibnamefont
  {Nelson}},\ }\href@noop {} {\bibfield  {journal} {\bibinfo  {journal}
  {Evolution}\ }\textbf {\bibinfo {volume} {64}},\ \bibinfo {pages} {193}
  (\bibinfo {year} {2009})}\BibitemShut {NoStop}%
\bibitem [{\citenamefont {Korolev}\ \emph {et~al.}(2012)\citenamefont
  {Korolev}, \citenamefont {M\"{u}ller}, \citenamefont {Karahan}, \citenamefont
  {Murray}, \citenamefont {Hallatschek},\ and\ \citenamefont
  {Nelson}}]{Korolev2012}%
  \BibitemOpen
  \bibfield  {author} {\bibinfo {author} {\bibfnamefont {K.~S.}\ \bibnamefont
  {Korolev}}, \bibinfo {author} {\bibfnamefont {M.~J.~I.}\ \bibnamefont
  {M\"{u}ller}}, \bibinfo {author} {\bibfnamefont {N.}~\bibnamefont {Karahan}},
  \bibinfo {author} {\bibfnamefont {A.~W.}\ \bibnamefont {Murray}}, \bibinfo
  {author} {\bibfnamefont {O.}~\bibnamefont {Hallatschek}}, \ and\ \bibinfo
  {author} {\bibfnamefont {D.~R.}\ \bibnamefont {Nelson}},\ }\href@noop {}
  {\bibfield  {journal} {\bibinfo  {journal} {Phys. Biol.}\ }\textbf {\bibinfo
  {volume} {9}},\ \bibinfo {pages} {026008} (\bibinfo {year}
  {2012})}\BibitemShut {NoStop}%
\bibitem [{\citenamefont {Farrell}\ \emph {et~al.}(2017)\citenamefont
  {Farrell}, \citenamefont {Gralka}, \citenamefont {Hallatschek},\ and\
  \citenamefont {Waclaw}}]{Farrell2017}%
  \BibitemOpen
  \bibfield  {author} {\bibinfo {author} {\bibfnamefont {F.~D.}\ \bibnamefont
  {Farrell}}, \bibinfo {author} {\bibfnamefont {M.}~\bibnamefont {Gralka}},
  \bibinfo {author} {\bibfnamefont {O.}~\bibnamefont {Hallatschek}}, \ and\
  \bibinfo {author} {\bibfnamefont {B.}~\bibnamefont {Waclaw}},\ }\href@noop {}
  {\bibfield  {journal} {\bibinfo  {journal} {J. R. Soc. Interface}\ }\textbf
  {\bibinfo {volume} {14}},\ \bibinfo {pages} {20170073} (\bibinfo {year}
  {2017})}\BibitemShut {NoStop}%
\bibitem [{\citenamefont {Saito}\ and\ \citenamefont
  {M\"{u}ller-Krumbhaar}(1995)}]{Saito1995}%
  \BibitemOpen
  \bibfield  {author} {\bibinfo {author} {\bibfnamefont {Y.}~\bibnamefont
  {Saito}}\ and\ \bibinfo {author} {\bibfnamefont {H.}~\bibnamefont
  {M\"{u}ller-Krumbhaar}},\ }\href@noop {} {\bibfield  {journal} {\bibinfo
  {journal} {Phys. Rev. Lett.}\ }\textbf {\bibinfo {volume} {74}},\ \bibinfo
  {pages} {4325 } (\bibinfo {year} {1995})}\BibitemShut {NoStop}%
\bibitem [{\citenamefont {Kuhr}\ \emph {et~al.}(2011)\citenamefont {Kuhr},
  \citenamefont {Leisner},\ and\ \citenamefont {Frey}}]{Kuhr2011}%
  \BibitemOpen
  \bibfield  {author} {\bibinfo {author} {\bibfnamefont {J.-T.}\ \bibnamefont
  {Kuhr}}, \bibinfo {author} {\bibfnamefont {M.}~\bibnamefont {Leisner}}, \
  and\ \bibinfo {author} {\bibfnamefont {E.}~\bibnamefont {Frey}},\ }\href@noop
  {} {\bibfield  {journal} {\bibinfo  {journal} {New J. Phys.}\ }\textbf
  {\bibinfo {volume} {13}},\ \bibinfo {pages} {113013} (\bibinfo {year}
  {2011})}\BibitemShut {NoStop}%
\bibitem [{\citenamefont {Lavrentovich}\ and\ \citenamefont
  {Nelson}(2014)}]{Lavrentovich2014}%
  \BibitemOpen
  \bibfield  {author} {\bibinfo {author} {\bibfnamefont {M.~O.}\ \bibnamefont
  {Lavrentovich}}\ and\ \bibinfo {author} {\bibfnamefont {D.~R.}\ \bibnamefont
  {Nelson}},\ }\href@noop {} {\bibfield  {journal} {\bibinfo  {journal} {Phys.
  Rev. Lett.}\ }\textbf {\bibinfo {volume} {112}},\ \bibinfo {pages} {138102}
  (\bibinfo {year} {2014})}\BibitemShut {NoStop}%
\bibitem [{\citenamefont {Kuhr}\ and\ \citenamefont {Stark}(2015)}]{Kuhr2015}%
  \BibitemOpen
  \bibfield  {author} {\bibinfo {author} {\bibfnamefont {J.-T.}\ \bibnamefont
  {Kuhr}}\ and\ \bibinfo {author} {\bibfnamefont {H.}~\bibnamefont {Stark}},\
  }\href@noop {} {\bibfield  {journal} {\bibinfo  {journal} {New J. Phys.}\
  }\textbf {\bibinfo {volume} {17}},\ \bibinfo {pages} {103035} (\bibinfo
  {year} {2015})}\BibitemShut {NoStop}%
\bibitem [{\citenamefont {Kardar}\ \emph {et~al.}(1986)\citenamefont {Kardar},
  \citenamefont {Parisi},\ and\ \citenamefont {Zhang}}]{Kardar1986}%
  \BibitemOpen
  \bibfield  {author} {\bibinfo {author} {\bibfnamefont {M.}~\bibnamefont
  {Kardar}}, \bibinfo {author} {\bibfnamefont {G.}~\bibnamefont {Parisi}}, \
  and\ \bibinfo {author} {\bibfnamefont {Y.-C.}\ \bibnamefont {Zhang}},\
  }\href@noop {} {\bibfield  {journal} {\bibinfo  {journal} {Phys. Rev. Lett.}\
  }\textbf {\bibinfo {volume} {56}},\ \bibinfo {pages} {889} (\bibinfo {year}
  {1986})}\BibitemShut {NoStop}%
\bibitem [{\citenamefont {Drossel}\ and\ \citenamefont
  {Kardar}(2000)}]{Drossel2000}%
  \BibitemOpen
  \bibfield  {author} {\bibinfo {author} {\bibfnamefont {B.}~\bibnamefont
  {Drossel}}\ and\ \bibinfo {author} {\bibfnamefont {M.}~\bibnamefont
  {Kardar}},\ }\href@noop {} {\bibfield  {journal} {\bibinfo  {journal} {Phys.
  Rev. Lett.}\ }\textbf {\bibinfo {volume} {85}},\ \bibinfo {pages} {614}
  (\bibinfo {year} {2000})}\BibitemShut {NoStop}%
\bibitem [{\citenamefont {Drossel}\ and\ \citenamefont
  {Kardar}(2003)}]{Drossel2003}%
  \BibitemOpen
  \bibfield  {author} {\bibinfo {author} {\bibfnamefont {B.}~\bibnamefont
  {Drossel}}\ and\ \bibinfo {author} {\bibfnamefont {M.}~\bibnamefont
  {Kardar}},\ }\href@noop {} {\bibfield  {journal} {\bibinfo  {journal} {Eur.
  Phys. J. B}\ }\textbf {\bibinfo {volume} {36}},\ \bibinfo {pages} {401}
  (\bibinfo {year} {2003})}\BibitemShut {NoStop}%
\bibitem [{\citenamefont {Ertas}\ and\ \citenamefont {Kardar}(1993)}]{EK93}%
  \BibitemOpen
  \bibfield  {author} {\bibinfo {author} {\bibfnamefont {D.}~\bibnamefont
  {Ertas}}\ and\ \bibinfo {author} {\bibfnamefont {M.}~\bibnamefont {Kardar}},\
  }\href {\doibase 10.1103/PhysRevE.48.1228} {\bibfield  {journal} {\bibinfo
  {journal} {Phys. Rev. E}\ }\textbf {\bibinfo {volume} {48}},\ \bibinfo
  {pages} {1228} (\bibinfo {year} {1993})}\BibitemShut {NoStop}%
\bibitem [{\citenamefont {Sarkar}\ and\ \citenamefont
  {Basu}(2012)}]{Sarkar2012}%
  \BibitemOpen
  \bibfield  {author} {\bibinfo {author} {\bibfnamefont {N.}~\bibnamefont
  {Sarkar}}\ and\ \bibinfo {author} {\bibfnamefont {A.}~\bibnamefont {Basu}},\
  }\href@noop {} {\bibfield  {journal} {\bibinfo  {journal} {Phys. Rev. E}\
  }\textbf {\bibinfo {volume} {86}},\ \bibinfo {pages} {021122} (\bibinfo
  {year} {2012})}\BibitemShut {NoStop}%
\bibitem [{\citenamefont {Kimura}\ and\ \citenamefont
  {Weiss}(1964)}]{Kimura1964}%
  \BibitemOpen
  \bibfield  {author} {\bibinfo {author} {\bibfnamefont {M.}~\bibnamefont
  {Kimura}}\ and\ \bibinfo {author} {\bibfnamefont {G.~H.}\ \bibnamefont
  {Weiss}},\ }\href@noop {} {\bibfield  {journal} {\bibinfo  {journal}
  {Genetics}\ }\textbf {\bibinfo {volume} {49}},\ \bibinfo {pages} {561}
  (\bibinfo {year} {1964})}\BibitemShut {NoStop}%
\bibitem [{\citenamefont {Lavrentovich}\ \emph {et~al.}(2013)\citenamefont
  {Lavrentovich}, \citenamefont {Korolev},\ and\ \citenamefont
  {Nelson}}]{Lavrentovich2013}%
  \BibitemOpen
  \bibfield  {author} {\bibinfo {author} {\bibfnamefont {M.~O.}\ \bibnamefont
  {Lavrentovich}}, \bibinfo {author} {\bibfnamefont {K.~S.}\ \bibnamefont
  {Korolev}}, \ and\ \bibinfo {author} {\bibfnamefont {D.~R.}\ \bibnamefont
  {Nelson}},\ }\href@noop {} {\bibfield  {journal} {\bibinfo  {journal} {Phys.
  Rev. E}\ }\textbf {\bibinfo {volume} {87}},\ \bibinfo {pages} {012103}
  (\bibinfo {year} {2013})}\BibitemShut {NoStop}%
\bibitem [{\citenamefont {Meakin}\ \emph {et~al.}(1986)\citenamefont {Meakin},
  \citenamefont {Ramanlal}, \citenamefont {Sander},\ and\ \citenamefont
  {Ball}}]{Meakin1986}%
  \BibitemOpen
  \bibfield  {author} {\bibinfo {author} {\bibfnamefont {P.}~\bibnamefont
  {Meakin}}, \bibinfo {author} {\bibfnamefont {P.}~\bibnamefont {Ramanlal}},
  \bibinfo {author} {\bibfnamefont {L.~M.}\ \bibnamefont {Sander}}, \ and\
  \bibinfo {author} {\bibfnamefont {R.~C.}\ \bibnamefont {Ball}},\ }\href@noop
  {} {\bibfield  {journal} {\bibinfo  {journal} {Phys. Rev. A}\ }\textbf
  {\bibinfo {volume} {34}},\ \bibinfo {pages} {5091} (\bibinfo {year}
  {1986})}\BibitemShut {NoStop}%
\bibitem [{\citenamefont {Plischke}\ \emph {et~al.}(1987)\citenamefont
  {Plischke}, \citenamefont {R{\'a}cz},\ and\ \citenamefont
  {Liu}}]{Plischke1987}%
  \BibitemOpen
  \bibfield  {author} {\bibinfo {author} {\bibfnamefont {M.}~\bibnamefont
  {Plischke}}, \bibinfo {author} {\bibfnamefont {Z.}~\bibnamefont {R{\'a}cz}},
  \ and\ \bibinfo {author} {\bibfnamefont {D.}~\bibnamefont {Liu}},\
  }\href@noop {} {\bibfield  {journal} {\bibinfo  {journal} {Phys. Rev. B}\
  }\textbf {\bibinfo {volume} {35}},\ \bibinfo {pages} {3485} (\bibinfo {year}
  {1987})}\BibitemShut {NoStop}%
\bibitem [{\citenamefont {Janssen}(2005)}]{Janssen2005b}%
  \BibitemOpen
  \bibfield  {author} {\bibinfo {author} {\bibfnamefont {H.-K.}\ \bibnamefont
  {Janssen}},\ }\href@noop {} {\bibfield  {journal} {\bibinfo  {journal} {J.
  Phys.: Condens. Matter}\ }\textbf {\bibinfo {volume} {17}},\ \bibinfo {pages}
  {S1973} (\bibinfo {year} {2005})}\BibitemShut {NoStop}%
\bibitem [{\citenamefont {Edwards}\ \emph {et~al.}(1982)\citenamefont
  {Edwards}, \citenamefont {F.R.S.},\ and\ \citenamefont {Wilkinson}}]{EW82}%
  \BibitemOpen
  \bibfield  {author} {\bibinfo {author} {\bibfnamefont {S.~F.}\ \bibnamefont
  {Edwards}}, \bibinfo {author} {\bibnamefont {F.R.S.}}, \ and\ \bibinfo
  {author} {\bibfnamefont {D.~R.}\ \bibnamefont {Wilkinson}},\ }\href@noop {}
  {\bibfield  {journal} {\bibinfo  {journal} {Proc. R. Soc. Lond. A}\ }\textbf
  {\bibinfo {volume} {381}},\ \bibinfo {pages} {17} (\bibinfo {year}
  {1982})}\BibitemShut {NoStop}%
\bibitem [{\citenamefont {Krug}\ \emph {et~al.}(1992)\citenamefont {Krug},
  \citenamefont {Meakin},\ and\ \citenamefont {Halpin-Healy}}]{Krug1992}%
  \BibitemOpen
  \bibfield  {author} {\bibinfo {author} {\bibfnamefont {J.}~\bibnamefont
  {Krug}}, \bibinfo {author} {\bibfnamefont {P.}~\bibnamefont {Meakin}}, \ and\
  \bibinfo {author} {\bibfnamefont {T.}~\bibnamefont {Halpin-Healy}},\
  }\href@noop {} {\bibfield  {journal} {\bibinfo  {journal} {Phys. Rev. A}\
  }\textbf {\bibinfo {volume} {45}},\ \bibinfo {pages} {638} (\bibinfo {year}
  {1992})}\BibitemShut {NoStop}%
\bibitem [{\citenamefont {Tang}\ \emph {et~al.}(1995)\citenamefont {Tang},
  \citenamefont {Kardar},\ and\ \citenamefont {Dhar}}]{TKDprl}%
  \BibitemOpen
  \bibfield  {author} {\bibinfo {author} {\bibfnamefont {L.-H.}\ \bibnamefont
  {Tang}}, \bibinfo {author} {\bibfnamefont {M.}~\bibnamefont {Kardar}}, \ and\
  \bibinfo {author} {\bibfnamefont {D.}~\bibnamefont {Dhar}},\ }\href@noop {}
  {\bibfield  {journal} {\bibinfo  {journal} {Phys. Rev. Lett.}\ }\textbf
  {\bibinfo {volume} {74}},\ \bibinfo {pages} {920} (\bibinfo {year}
  {1995})}\BibitemShut {NoStop}%
\bibitem [{Note1()}]{Note1}%
  \BibitemOpen
  \bibinfo {note} {For $\lambda =0$, the bulges and dips will have parabolic
  shapes.}\BibitemShut {Stop}%
\bibitem [{\citenamefont {Eigen}(1971)}]{Eigen1971}%
  \BibitemOpen
  \bibfield  {author} {\bibinfo {author} {\bibfnamefont {M.}~\bibnamefont
  {Eigen}},\ }\href {\doibase 10.1007/BF00623322} {\bibfield  {journal}
  {\bibinfo  {journal} {Naturwissenschaften}\ }\textbf {\bibinfo {volume}
  {58}},\ \bibinfo {pages} {465} (\bibinfo {year} {1971})}\BibitemShut
  {NoStop}%
\bibitem [{\citenamefont {Eigen}\ and\ \citenamefont
  {Schuster}(1977)}]{Eigen1977}%
  \BibitemOpen
  \bibfield  {author} {\bibinfo {author} {\bibfnamefont {M.}~\bibnamefont
  {Eigen}}\ and\ \bibinfo {author} {\bibfnamefont {P.}~\bibnamefont
  {Schuster}},\ }\href@noop {} {\bibfield  {journal} {\bibinfo  {journal}
  {Naturwissenschaften}\ }\textbf {\bibinfo {volume} {64}},\ \bibinfo {pages}
  {541} (\bibinfo {year} {1977})}\BibitemShut {NoStop}%
\bibitem [{\citenamefont {Eigen}\ \emph {et~al.}(1988)\citenamefont {Eigen},
  \citenamefont {McCaskill},\ and\ \citenamefont {Schuster}}]{Eigen1988}%
  \BibitemOpen
  \bibfield  {author} {\bibinfo {author} {\bibfnamefont {M.}~\bibnamefont
  {Eigen}}, \bibinfo {author} {\bibfnamefont {J.}~\bibnamefont {McCaskill}}, \
  and\ \bibinfo {author} {\bibfnamefont {P.}~\bibnamefont {Schuster}},\
  }\href@noop {} {\bibfield  {journal} {\bibinfo  {journal} {J. Phys. Chem.}\
  }\textbf {\bibinfo {volume} {92}},\ \bibinfo {pages} {6881} (\bibinfo {year}
  {1988})}\BibitemShut {NoStop}%
\bibitem [{\citenamefont {Eigen}\ \emph {et~al.}(1989)\citenamefont {Eigen},
  \citenamefont {McCaskill},\ and\ \citenamefont {Schuster}}]{Eigen1989}%
  \BibitemOpen
  \bibfield  {author} {\bibinfo {author} {\bibfnamefont {M.}~\bibnamefont
  {Eigen}}, \bibinfo {author} {\bibfnamefont {J.}~\bibnamefont {McCaskill}}, \
  and\ \bibinfo {author} {\bibfnamefont {P.}~\bibnamefont {Schuster}},\
  }\href@noop {} {\bibfield  {journal} {\bibinfo  {journal} {Adv. Chem. Phys.}\
  }\textbf {\bibinfo {volume} {75}},\ \bibinfo {pages} {149} (\bibinfo {year}
  {1989})}\BibitemShut {NoStop}%
\bibitem [{\citenamefont {Leuth\:ausser}(1986)}]{Leuthausser1986}%
  \BibitemOpen
  \bibfield  {author} {\bibinfo {author} {\bibfnamefont {I.}~\bibnamefont
  {Leuth\:ausser}},\ }\href@noop {} {\bibfield  {journal} {\bibinfo  {journal}
  {J. Chem. Phys.}\ }\textbf {\bibinfo {volume} {84}},\ \bibinfo {pages} {1884}
  (\bibinfo {year} {1986})}\BibitemShut {NoStop}%
\bibitem [{\citenamefont {Leuth\:ausser}(1987)}]{Leuthausser1987}%
  \BibitemOpen
  \bibfield  {author} {\bibinfo {author} {\bibfnamefont {I.}~\bibnamefont
  {Leuth\:ausser}},\ }\href@noop {} {\bibfield  {journal} {\bibinfo  {journal}
  {J. Stat. Phys.}\ }\textbf {\bibinfo {volume} {48}},\ \bibinfo {pages} {343}
  (\bibinfo {year} {1987})}\BibitemShut {NoStop}%
\bibitem [{\citenamefont {Domingo}\ and\ \citenamefont
  {Schuster}(2016)}]{Quasispecies}%
  \BibitemOpen
  \bibinfo {editor} {\bibfnamefont {E.}~\bibnamefont {Domingo}}\ and\ \bibinfo
  {editor} {\bibfnamefont {P.}~\bibnamefont {Schuster}},\ eds.,\ \href@noop {}
  {\emph {\bibinfo {title} {Quasispcies: From Theory to Experimental
  Systems}}},\ \bibinfo {series} {Current Topics in Microbiology and
  Immunology}, Vol.\ \bibinfo {volume} {392}\ (\bibinfo  {publisher} {Springer,
  Swizterland},\ \bibinfo {year} {2016})\BibitemShut {NoStop}%
\bibitem [{\citenamefont {Kardar}(2007)}]{Kardar}%
  \BibitemOpen
  \bibfield  {author} {\bibinfo {author} {\bibfnamefont {M.}~\bibnamefont
  {Kardar}},\ }\href@noop {} {\emph {\bibinfo {title} {Statistical physics of
  fields}}}\ (\bibinfo  {publisher} {Cambridge University Press, UK},\ \bibinfo
  {year} {2007})\BibitemShut {NoStop}%
\bibitem [{\citenamefont {Kardar}(1999)}]{Kardar1999}%
  \BibitemOpen
  \bibfield  {author} {\bibinfo {author} {\bibfnamefont {M.}~\bibnamefont
  {Kardar}},\ }\href@noop {} {\bibfield  {journal} {\bibinfo  {journal}
  {Physica A}\ }\textbf {\bibinfo {volume} {263}},\ \bibinfo {pages} {345}
  (\bibinfo {year} {1999})}\BibitemShut {NoStop}%
\bibitem [{\citenamefont {Halpin-Healy}\ and\ \citenamefont
  {Takeuchi}(2015)}]{H-HK15}%
  \BibitemOpen
  \bibfield  {author} {\bibinfo {author} {\bibfnamefont {T.}~\bibnamefont
  {Halpin-Healy}}\ and\ \bibinfo {author} {\bibfnamefont {K.~A.}\ \bibnamefont
  {Takeuchi}},\ }\href {\doibase 10.1007/s10955-015-1282-1} {\bibfield
  {journal} {\bibinfo  {journal} {Journal of Statistical Physics}\ }\textbf
  {\bibinfo {volume} {160}},\ \bibinfo {pages} {794} (\bibinfo {year}
  {2015})}\BibitemShut {NoStop}%
\bibitem [{\citenamefont {Frey}\ and\ \citenamefont
  {T\"{a}uber}(1994)}]{Frey1994}%
  \BibitemOpen
  \bibfield  {author} {\bibinfo {author} {\bibfnamefont {E.}~\bibnamefont
  {Frey}}\ and\ \bibinfo {author} {\bibfnamefont {U.~C.}\ \bibnamefont
  {T\"{a}uber}},\ }\href@noop {} {\bibfield  {journal} {\bibinfo  {journal}
  {Phys. Rev. E}\ }\textbf {\bibinfo {volume} {50}},\ \bibinfo {pages} {1024}
  (\bibinfo {year} {1994})}\BibitemShut {NoStop}%
\bibitem [{\citenamefont {Burgers}(1974)}]{BurgersBook}%
  \BibitemOpen
  \bibfield  {author} {\bibinfo {author} {\bibfnamefont {J.~M.}\ \bibnamefont
  {Burgers}},\ }\href@noop {} {\emph {\bibinfo {title} {The nonlinear diffusion
  equation : asymptotic solutions and statistical problems}}}\ (\bibinfo {year}
  {1974})\BibitemShut {NoStop}%
\bibitem [{\citenamefont {Jensen}(1996)}]{Jensen1996}%
  \BibitemOpen
  \bibfield  {author} {\bibinfo {author} {\bibfnamefont {I.}~\bibnamefont
  {Jensen}},\ }\href@noop {} {\bibfield  {journal} {\bibinfo  {journal} {J.
  Phys. A Math. Gen.}\ }\textbf {\bibinfo {volume} {29}},\ \bibinfo {pages}
  {7013} (\bibinfo {year} {1996})}\BibitemShut {NoStop}%
\bibitem [{\citenamefont {T\"{a}uber}(2012)}]{Tauber2012}%
  \BibitemOpen
  \bibfield  {author} {\bibinfo {author} {\bibfnamefont {U.~C.}\ \bibnamefont
  {T\"{a}uber}},\ }\href@noop {} {\bibfield  {journal} {\bibinfo  {journal}
  {Nucl. Phys. B-Proc. Sup.}\ }\textbf {\bibinfo {volume} {228}},\ \bibinfo
  {pages} {7} (\bibinfo {year} {2012})}\BibitemShut {NoStop}%
\bibitem [{\citenamefont {Medina}\ \emph {et~al.}(1989)\citenamefont {Medina},
  \citenamefont {Hwa}, \citenamefont {Kardar},\ and\ \citenamefont
  {Zhang}}]{Medina1989}%
  \BibitemOpen
  \bibfield  {author} {\bibinfo {author} {\bibfnamefont {E.}~\bibnamefont
  {Medina}}, \bibinfo {author} {\bibfnamefont {T.}~\bibnamefont {Hwa}},
  \bibinfo {author} {\bibfnamefont {M.}~\bibnamefont {Kardar}}, \ and\ \bibinfo
  {author} {\bibfnamefont {Y.-C.}\ \bibnamefont {Zhang}},\ }\href@noop {}
  {\bibfield  {journal} {\bibinfo  {journal} {Phys. Rev. A}\ }\textbf {\bibinfo
  {volume} {39}},\ \bibinfo {pages} {3053} (\bibinfo {year}
  {1989})}\BibitemShut {NoStop}%
\bibitem [{\citenamefont {Folk}\ and\ \citenamefont {Moser}(2006)}]{Folk2006}%
  \BibitemOpen
  \bibfield  {author} {\bibinfo {author} {\bibfnamefont {R.}~\bibnamefont
  {Folk}}\ and\ \bibinfo {author} {\bibfnamefont {G.}~\bibnamefont {Moser}},\
  }\href@noop {} {\bibfield  {journal} {\bibinfo  {journal} {J. Phys. A}\
  }\textbf {\bibinfo {volume} {39}},\ \bibinfo {pages} {R207} (\bibinfo {year}
  {2006})}\BibitemShut {NoStop}%
\bibitem [{\citenamefont {Akkineni}\ and\ \citenamefont
  {T\"{a}uber}(2004)}]{Akkineni2004}%
  \BibitemOpen
  \bibfield  {author} {\bibinfo {author} {\bibfnamefont {V.~K.}\ \bibnamefont
  {Akkineni}}\ and\ \bibinfo {author} {\bibfnamefont {U.~C.}\ \bibnamefont
  {T\"{a}uber}},\ }\href@noop {} {\bibfield  {journal} {\bibinfo  {journal}
  {Phys. Rev. E}\ }\textbf {\bibinfo {volume} {69}},\ \bibinfo {pages} {036113}
  (\bibinfo {year} {2004})}\BibitemShut {NoStop}%
\bibitem [{\citenamefont {Das}\ \emph {et~al.}(2001)\citenamefont {Das},
  \citenamefont {Basu}, \citenamefont {Barma},\ and\ \citenamefont
  {Ramaswamy}}]{Das2001}%
  \BibitemOpen
  \bibfield  {author} {\bibinfo {author} {\bibfnamefont {D.}~\bibnamefont
  {Das}}, \bibinfo {author} {\bibfnamefont {A.}~\bibnamefont {Basu}}, \bibinfo
  {author} {\bibfnamefont {M.}~\bibnamefont {Barma}}, \ and\ \bibinfo {author}
  {\bibfnamefont {S.}~\bibnamefont {Ramaswamy}},\ }\href@noop {} {\bibfield
  {journal} {\bibinfo  {journal} {Phys. Rev. E}\ }\textbf {\bibinfo {volume}
  {64}},\ \bibinfo {pages} {021402} (\bibinfo {year} {2001})}\BibitemShut
  {NoStop}%
\bibitem [{\citenamefont {Sarkar}\ and\ \citenamefont
  {Basu}(2014)}]{Sarkar2014}%
  \BibitemOpen
  \bibfield  {author} {\bibinfo {author} {\bibfnamefont {N.}~\bibnamefont
  {Sarkar}}\ and\ \bibinfo {author} {\bibfnamefont {A.}~\bibnamefont {Basu}},\
  }\href@noop {} {\bibfield  {journal} {\bibinfo  {journal} {J. Stat. Mech.}\
  ,\ \bibinfo {pages} {P08016}} (\bibinfo {year} {2014})}\BibitemShut {NoStop}%
\bibitem [{\citenamefont {Banerjee}\ and\ \citenamefont
  {Basu}(2018)}]{Banerjee2018}%
  \BibitemOpen
  \bibfield  {author} {\bibinfo {author} {\bibfnamefont {T.}~\bibnamefont
  {Banerjee}}\ and\ \bibinfo {author} {\bibfnamefont {A.}~\bibnamefont
  {Basu}},\ }\href@noop {} {\bibfield  {journal} {\bibinfo  {journal} {J. Stat.
  Mech.}\ ,\ \bibinfo {pages} {013202}} (\bibinfo {year} {2018})}\BibitemShut
  {NoStop}%
\bibitem [{Note2()}]{Note2}%
  \BibitemOpen
  \bibinfo {note} {There is one fixed point for which all nonlinear couplings
  are stable and $\theta $ is marginally stable ($d\theta /dl|_{\theta =0}=0$),
  but flows $\theta \to \infty $ for any finite value of $\theta $: $u=0$,
  $\zeta =2\epsilon $, $\delta _1=-2\epsilon $, and $\delta _2=0$. To verify
  the stability of this fixed point, higher order terms need to be
  considered.}\BibitemShut {Stop}%
\bibitem [{\citenamefont {George}\ and\ \citenamefont
  {Korolev}()}]{George2017}%
  \BibitemOpen
  \bibfield  {author} {\bibinfo {author} {\bibfnamefont {A.~B.}\ \bibnamefont
  {George}}\ and\ \bibinfo {author} {\bibfnamefont {K.~S.}\ \bibnamefont
  {Korolev}},\ }\href@noop {} {\enquote {\bibinfo {title} {Chirality provides a
  direct fitness advantage and facilitates intermixing cellular aggregates},}\
  }\bibinfo {note} {Arxiv:1710.09459}\BibitemShut {NoStop}%
\bibitem [{\citenamefont {Plummer}\ \emph {et~al.}()\citenamefont {Plummer},
  \citenamefont {Benzi}, \citenamefont {Nelson},\ and\ \citenamefont
  {Toschi}}]{Plummer2018}%
  \BibitemOpen
  \bibfield  {author} {\bibinfo {author} {\bibfnamefont {A.}~\bibnamefont
  {Plummer}}, \bibinfo {author} {\bibfnamefont {R.}~\bibnamefont {Benzi}},
  \bibinfo {author} {\bibfnamefont {D.~R.}\ \bibnamefont {Nelson}}, \ and\
  \bibinfo {author} {\bibfnamefont {F.}~\bibnamefont {Toschi}},\ }\href@noop {}
  {\enquote {\bibinfo {title} {Fixation probabilities in weakly compressible
  fluid flows},}\ }\bibinfo {note} {Arxiv:1808.07128}\BibitemShut {NoStop}%
\bibitem [{\citenamefont {Antonov}\ \emph {et~al.}(2009)\citenamefont
  {Antonov}, \citenamefont {Iglovikov},\ and\ \citenamefont
  {Kapustin}}]{Antonov2009}%
  \BibitemOpen
  \bibfield  {author} {\bibinfo {author} {\bibfnamefont {N.}~\bibnamefont
  {Antonov}}, \bibinfo {author} {\bibfnamefont {V.}~\bibnamefont {Iglovikov}},
  \ and\ \bibinfo {author} {\bibfnamefont {A.}~\bibnamefont {Kapustin}},\
  }\href@noop {} {\bibfield  {journal} {\bibinfo  {journal} {J. Phys. A Math.
  Theor.}\ }\textbf {\bibinfo {volume} {42}},\ \bibinfo {pages} {135001}
  (\bibinfo {year} {2009})}\BibitemShut {NoStop}%
\end{thebibliography}%
\end{document}